\documentclass[11pt]{article}

\usepackage{amsmath}
\usepackage{graphicx}
\usepackage{indentfirst}
\usepackage{amssymb}
\usepackage{cite}
\usepackage{color}
\usepackage{subfigure}
\usepackage{varwidth}
\usepackage{subfigure}
\usepackage[colorlinks=true, linkcolor=red, citecolor=blue, urlcolor=magenta]{hyperref}
\usepackage{xcolor}
\usepackage{amsmath}
\usepackage{bm}
\usepackage[normalem]{ulem}

\begin{document}

\title{Shadow and Polarization Images of Rotating Black Holes in Kalb-Ramond Gravity Illuminated by Several Thick Accretion Disks}

\date{}
\maketitle

\begin{center}
\author{Chen-Yu Yang,}$^{a}$\footnote{E-mail: chenyu\_yang2024@163.com}
\author{Huan Ye,}$^{b}$\footnote{E-mail: yehuan10053@163.com}
\author{Xiao-Xiong Zeng}$^{c}$\footnote{E-mail: xxzengphysics@163.com (Corresponding author)}
\\

\vskip 0.25in
$^{a}$\it{Department of Mechanics, Chongqing Jiaotong University, Chongqing 400000, People's Republic of China}\\
$^{b}$\it{School of Material Science and Engineering, Chongqing Jiaotong University, Chongqing 400074, People's Republic of China}\\
$^{c}$\it{College of Physics and Electronic Engineering, Chongqing Normal University, Chongqing 401331, People's Republic of China}\\

\end{center}

\vskip 0.6in
{\abstract{
Using ray-tracing techniques, this paper investigates the optical and polarization images of rotating black holes in Kalb-Ramond (KR) gravity illuminated by thick accretion disks. We examine two accretion disk models: the phenomenological radiatively inefficient accretion flow (RIAF) model and the analytical ballistic approximation accretion flow (BAAF) model. The RIAF model incorporates both isotropic and anisotropic radiation. In all models, the external bright rings corresponding to the high-order image and  the internal dark region  associated with the event horizon are observed. At high observational inclinations, the inner shadows are obscured by the radiation from the equatorial plane, which is significantly different from the thin accretion disk model. The primary distinction between isotropic and anisotropic radiation is that the latter causes distortion of the high-order image in the vertical direction, resulting in an elliptical structure. For the BAAF model, due to certain regions are geometrically thinner under the conical approximation,  the high-order images are  narrower compared to the RIAF model. Furthermore, we find that an increase in the rotational parameter $a$ leads to an asymmetry in the intensity distribution of the high-order image, while an increase in the spontaneous Lorentz violating parameters, $\varsigma$ and $\varpi$, results in a decrease in the size of the high-order image. In the polarization image, the  linear polarization is found to be  significantly influenced by the intensity, while it is relatively less affected by the parameters $\varsigma$ and $\varpi$. The electric vector position angle is mainly affected by the parameters $\varsigma$ and $\varpi$.
}}

\thispagestyle{empty}
\newpage
\setcounter{page}{1}


\section{Introduction}

In the past decade, the successful detection of gravitational waves~\cite{abbott2016observation,abbott2016gw151226} and the successful observation of supermassive black holes M87* and Sgr A*~\cite{event2019first,akiyama2022first} have injected new vitality into gravitational research, marking a new stage in the exploration of the nature of gravity. In black hole images, the center is a dark region surrounded by a bright, asymmetric photon ring. The specific features of the bright region are related not only to the black hole itself but also to its surrounding   plasma environment~\cite{gralla2019black}. Therefore, in different gravitational spacetimes, establishing theoretical models of black hole shadow  that align with observational images after considering various matter fields has become one of the key topics in current astrophysical research~\cite{yang2025observational,yang2025shadow}.

Supermassive black holes accrete hot, magnetized plasma, forming a bright accretion disk. The frequency observed in the image is dominated by thermal synchrotron radiation~\cite{lu2023ring,akiyama2019first}. Meanwhile, for high-spin black holes, electromagnetic energy may trigger relativistic jets~\cite{blandford1977electromagnetic}, known as "funnel wall jets" (FWJ). Theoretically, both the accretion disk and the jets can act as light sources, generating black hole shadow images. The study of black hole shadows has a long history~\cite{tsukamoto2014constraining,tsukamoto2018black,battista2024quantum,wang2025dynamical}, including spherically symmetric accretion models~\cite{narayan2019shadow,zeng2020shadows,heydari2023shadows}, optical and geometric thin accretion disk models~\cite{guo2024image,he2024observational,zeng2022shadows,li2021shadows}, and holographic Einstein rings~\cite{hashimoto2020imaging,aslam2024holographic,zeng2025schwarzschild}. Based on EHT observational data, many studies have also attempted to constrain the parameters of black hole models~\cite{zeng2025holographic,cui2024optical,guo2024influence,hou2024unique,huang2024images}. Current studies on black hole shadows mostly assume the accretion disk to be optically and geometrically thin. However, EHT and related observational studies suggest that in strong gravitational fields, the accretion flow near supermassive black holes, due to suppressed vertical cooling and matter compression, may become geometrically thick and optically thin~\cite{event2019first,akiyama2019first,akiyama2022first,ho1999spectral,narayan1994advection}. In this case, parameters such as electron number density, electron temperature, and magnetic field structure must be considered. In theoretical studies of geometrically thick accretion disks, most  work uses the radiatively inefficient accretion flow (RIAF) model with relatively low radiation efficiency, where the vertically-averaged electron number density and temperature typically follow a power-law distribution with radius~\cite{yuan2003nonthermal,broderick2005frequency,broderick2009imaging,wielgus2025semi,jiang2024shadows}.

Polarization images are important tools for probing the dynamics of matter and magnetic field structure around black holes~\cite{guo2024influence}. The polarization images of M87* and Sgr A* released by EHT reveal significant polarization distribution features within the emission rings, where the linear polarization images display a unique structure with the electric vector arranged in a spiral pattern~\cite{akiyama2021first,akiyama2024first}. To obtain information from polarization images, it is necessary to solve the null geodesic equation and the radiative transfer equation around the black hole. Based on this, the EHT collaboration reproduced the distribution of the electric vector position angle and the relative polarization intensity in the M87* polarization image. With the light-ray approximation derived by Beloborodov in the Schwarzschild black hole background~\cite{beloborodov2002gravitational,narayan2021polarized}, Gelles et al. constructed a simplified model with an equatorial emission source and generated the corresponding polarization image to the Kerr black hole~\cite{gelles2021polarized}. These studies suggest that the polarization characteristics are determined by the magnetic field geometry, the intrinsic parameters of the black hole, and the observation angle. In addition, polarization images of other types of black hole and horizonless ultra-compact objects have been extensively studied~\cite{yin2025bright,qin2022polarized,shi2024polarized,deliyski2023polarized,zeng2025polarization}, providing an effective approach to studying the accretion physics and potential spacetime geometric structures.

Advances in experimental techniques have not only enabled us to verify the applicability of General Relativity (GR) with higher precision and resolution~\cite{abbott2020gw190412,psaltis2020gravitational}, but have also driven the process of modifying GR and exploring alternative gravitational theories. An important method for modifying the GR is to alter the Einstein action. The introduction of the Kalb-Ramond (KR) field  is a key approach~\cite{kalb1974classical}. The KR field, as a two-form quantum field, is believed to be related to the closed string excitation in heterotic string theory~\cite{dine1985superstring}. Studies in the KR field have led to many important conclusions, such as the derivation of a third-order antisymmetric tensor, which is interpreted as the source of spacetime torsion and the intrinsic angular momentum of celestial bodies~\cite{majumdar1999parity,letelier1995spinning}. Further research on the gravitational effects of the KR field and its influence on particles can be found in~\cite{aashish2018inflation,chakraborty2017strong,kumar2020gravitational,aashish2019quantum,araujo2025particle1,araujo2025particle2,araujo2025non}.

Inspired by these groundbreaking studies, this paper intends to investigate the image characteristics of rotating black holes in KR gravity under illuminations of  two thick accretion disk models. On the one hand, we aim to study the impact of the thick accretion disk on the optical image of the black hole, particularly how the optical image differs under the illumination of different thick disks. On the other hand, we intend to analyze the influence of spontaneous Lorentz violating parameters, $\varsigma$ and $\varpi$, on the shadow image. In particular, we will examine whether this influence follows the same pattern as in the thin disk case. Furthermore, we will also investigate the polarization characteristics of an anisotropic thick accretion disk and compare them with the thin disk case.

The structure of this paper is organized as follows. Section~\ref{sec2} briefly introduces rotating black holes in KR gravity and presents the null geodesic equation near the black hole. Sections~\ref{sec3} and~\ref{sec4} introduce the electron radiation model and the accretion disk model, respectively. The electron radiation model incorporates both isotropic and anisotropic radiation, while the accretion disk model includes the phenomenological radiatively inefficient accretion flow (RIAF) model and the analytical ballistic approximation accretion flow (BAAF) model. Section~\ref{sec5} presents the basic theoretical background for polarization imaging in the case of anisotropic radiation. Section~\ref{sec6} provides the numerical results. Finally, Section~\ref{sec7} concludes with a summary and discussion. Unless otherwise specified, we adopt geometric units, i.e., $c = G = 1$, where $c$ is the vacuum speed of light and $G$ is the gravitational constant.

\section{Review of Black Holes in Kalb-Ramond Gravity}\label{sec2}
The rotating black hole solution in KR gravity is given by~\cite{kumar2020gravitational}
\begin{align}
	ds^{2}=&-\left(\frac{\Delta -a^{2}\sin^{2}\theta}{\rho^{2}}\right)dt^{2}+\frac{\rho^{2}}{\Delta }dr^{2}+\rho^{2}d\theta^{2}\nonumber\\
	&+\frac{\sin^{2}\theta}{\rho^{2}}\left[\left(r^{2}+a^{2}\right)^{2}-\Delta a^{2}\sin^{2}\theta\right]d\phi^{2}+\frac{2a\sin^{2}\theta}{\rho^{2}}\biggl(\Delta -a^{2}-r^{2}\biggr)dtd\phi,\label{eq:rkrmetric}
\end{align}
where
\begin{align}
	\Delta &=r^{2}-2Mr+a^{2}+\varsigma r^{\frac{2(\varpi-1)}{\varpi}},\label{Delta}\\
	\rho^{2}&=r^{2}+a^{2}\cos^{2}\theta,
\end{align}
in which  $M$ and $a$ are the mass and spin parameter of the black hole, and $\varsigma$ and $\varpi$ are the spontaneous Lorentz violating parameters, which are related to the vacuum expectation value and non-minimal coupling parameters of the KR field. When $\varpi \to 0$, the metric (\ref{eq:rkrmetric}) reduces to the Kerr black hole solution; when $\varpi = 1$, it reduces to the Kerr-Newman black hole solution; and when $a = 0$, it reduces to the static spherically symmetric black hole. In fact, the rotating KR black hole spacetime metric (\ref{eq:rkrmetric}) possesses time-translational and rotational invariance isometries, meaning that there exist two Killing vector fields $\left(\frac{\partial}{\partial t}\right)^{\mu}$ and $\left(\frac{\partial}{\partial \phi}\right)^{\mu}$.

The event horizon can be obtained by solving $\Delta=0$, i.e.,
\begin{equation}
	r^{2}-2Mr+a^{2}+\varsigma r^{\frac{2(\varpi-1)}{\varpi}}=0.
\end{equation}
For the special case $\varpi=1$, the above equation simplifies to the Kerr-Newman black hole
\begin{equation}
	r^2+a^2-2Mr+Q^2=0,
\end{equation}
where the spontaneous Lorentz violating parameter $\varsigma$ is replaced by $Q^2$. A zero angular momentum observer (ZAMO) is a static observer whose angular momentum is zero relative to spatial infinity. However, due to frame dragging, a ZAMO has a position-dependent angular velocity $\omega$. For the spacetime (\ref{eq:rkrmetric}), its value is
\begin{equation}
	\omega=\frac{d\phi}{dt}=-\frac{g_{t\phi}}{g_{\phi\phi}}=\frac{2Mar-\frac{\varsigma a}{r^{-2(\varpi-1)/\varpi}}}{(r^{2}+a^{2})^{2}-a^{2}\sin^{2}\theta\Delta}.
\end{equation}
As the observer approaches to the black hole, $\omega$ increases and reaches its maximum value at the event horizon
\begin{equation}
	\Omega=\omega\big|_{r=r_h} = 
	\frac{ 2 M a r_h - \frac{\varsigma a}{r_h^{-2(\varpi-1)/\varpi}} }{ (r_h^2 + a^2)^2 },
\end{equation}
where $r_h$ is the outer event horizon and $\Omega$ is the black hole's angular velocity. At this point, the observer is in corotation with the black hole. In the limit as $\varpi \to 0$, the above equation becomes
\begin{equation}
	\Omega=\frac{a}{r_h^2+a^2},
\end{equation}
which is consistent with the angular velocity of the Kerr black hole~\cite{chandrasekhar1998mathematical}.

We now provide a brief introduction to the photon motion near the rotating black hole in KR gravity. The null geodesic equations for the spacetime (\ref{eq:rkrmetric}) are given by~\cite{zubair2023rotating}
\begin{align}
	&\rho^{2}\dot{t}=a\big(\mathcal{L}-a\mathcal{E}\sin^{2}\theta\big)+\frac{r^{2}+a^{2}}{\Delta}\left[\mathcal{E}\big(r^{2}+a^{2}\big)-a\mathcal{L}\right], \\ 
	&\rho^{4}\dot{r}^2=\bm{R}(r), \\ 
	&\rho^{4}\dot{\theta}^2=\bm{\Theta}(\theta), \\ 
	&\rho^{2}\dot{\phi}=\left(\mathcal{L}\csc^{2}\theta-a\mathcal{E}\right)-\frac{a}{\Delta}\left[a\mathcal{L}-\mathcal{E}\big(r^{2}+a^{2}\big)\right],
\end{align}
where
\begin{align}
	&\bm{R}(r)=\left[\mathcal{E}(r^2+a^2)-a\mathcal{L}\right]^2-\Delta\left[\mathcal{K}+(\mathcal{L}-a\mathcal{E})^2\right], \\ 
	&\bm{\Theta}(\theta)=\mathcal{K}+\cos^2\theta\big(a^2\mathcal{E}^2-\mathcal{L}^2\csc^2\theta\big),
\end{align}
are the radial and angular potentials, respectively. The symbol "$\cdot{}$" represents the derivative with respect to the affine parameter, $\mathcal{E}$ and $\mathcal{L}$ are the two constants of motion obtained from the Killing vector fields, and $\mathcal{K}$ is the Carter constant. For convenience in subsequent calculations, we define the dimensionless impact parameters for photons as
\begin{equation}
	\xi\equiv\frac{\mathcal{L}}{\mathcal{E}},\quad \eta\equiv\frac{\mathcal{K}}{\mathcal{E}^2}.
\end{equation}

The boundary of the black hole shadow is surrounded by a luminous  sphere, known as the photon sphere~\cite{luminet1979image,synge1966escape,cunningham1972optical}. Photons that enter the photon sphere will fall into the event horizon, and only photons outside the photon sphere have a chance of reaching the observer. The radius of the photon sphere, $r_p$, can be determined by the radial potential and its derivative
\begin{equation}
	\bm{R}(r)\big|_{r=r_p} = 0, \qquad \left.\frac{\partial \bm{R}(r)}{\partial r}\right|_{r=r_p} = 0.\label{eq:rp1}
\end{equation}
Photons captured by the black hole, if they are in a stable orbit, remain bound indefinitely; however, if they are in an unstable orbit, they may escape after several orbits around the black hole. An unstable orbit must satisfy the following condition
\begin{equation}
	\left.\frac{\partial^2\bm{R}(r)}{\partial r^2}\right|_{r=r_p}>0.\label{eq:rp2}
\end{equation}
Solving the system of equations (\ref{eq:rp1}) and (\ref{eq:rp2}) gives a set of critical impact parameters corresponding to the unstable photon orbits
\begin{align}
	&\xi_{c} = \frac{\left(r_{p}^{2}+a^{2}\right)\Delta'(r_{p})-4r_{p}\Delta(r_{p})}{a\Delta'(r_{p})},\\
	&\eta_{c} = \frac{r_{p}^{2}\left[-16\Delta(r_{p})^{2}-r_{p}^{2}\Delta'(r_{p})^{2}+8\Delta(r_{p})\Big(2a^{2}+r_{p}\Delta'(r_{p})\Big)\right]}{a^{2}\Delta'(r_{p})^{2}},
\end{align}
where $\Delta^{\prime}$ represents the derivative with respect to the radial coordinate $r$.

\section{Electron Radiation Model}\label{sec3}
For the thick accretion disk model, parameters such as particle number density, electron temperature, and magnetic field structure are crucial. In theory, these physical quantities can be obtained by solving the general relativistic magnetohydrodynamic (GRMHD) equations. However, because of the high analytical complexity of the GRMHD equations, this paper will employ numerical methods and make appropriate simplifications.

\subsection{Radiative Transfer Equation}
First, we discuss the non-polarized case. The radiative transfer equation is given by
\begin{equation}
	\frac{d}{d\lambda}\tilde{I}=\tilde{J}-\tilde{\alpha}\tilde{I},\label{eq:rte}
\end{equation}
where $\tilde{I}, \tilde{J}, \tilde{\alpha}$ are generalized invariants, and their relations to the actual physical quantities are
\begin{equation}
	\tilde{I}=\frac{I_{\nu}}{\nu^{3}},\quad\tilde{J}=\frac{j_{\nu}}{\nu^{2}},\quad\tilde{\alpha}=\nu\alpha_{\nu},\label{eq:rcs}
\end{equation}
where $\nu$ is the photon frequency in the local reference frame, and $I_{\nu}, j_{\nu}, \alpha_{\nu}$ are the specific intensity, emissivity, and absorption coefficient, respectively. The solution to equation (\ref{eq:rte}) in geometric units is
\begin{equation}
	\tilde{I}(\lambda)=\tilde{I}(\lambda_0)+\int_{\lambda_0}^\lambda d\lambda^{\prime}\tilde{J}(\lambda^{\prime})\exp\left(-\int_{\lambda^{\prime}}^\lambda d\lambda^{\prime\prime}\tilde{\alpha}(\lambda^{\prime\prime})\right).
\end{equation}
To convert to CGS units, we multiply the affine parameter in equation (\ref{eq:rte}) by a factor $\mathcal{C}={r_{g}}/{\nu_{0}}$, where $r_{g}=GM/c^{2}$ is the unit length and $\nu_{0}$ is the true photon frequency at infinity. Equation (\ref{eq:rte}) becomes
\begin{equation}
	\frac1{\mathcal{C}}\frac{ d}{ d\lambda}\tilde{I}=\tilde{J}-\tilde{\alpha}\tilde{I},
\end{equation}
and its solution is
\begin{equation}
	I_{\nu}=\chi^{3}I_{\nu_{0}}+r_{g}\int_{\lambda_{0}}^{\lambda} d\lambda^{\prime}\chi^{2}j_{\nu}(\lambda^{\prime})\exp\left(-r_{g}\int_{\lambda^{\prime}}^{\lambda} d\lambda^{\prime\prime}\alpha_{\nu}(\lambda^{\prime\prime})/\chi\right),\label{eq:is}
\end{equation}
where $\chi=\nu_{0}/\nu$ is the redshift factor, $\nu$ is the frequency in the local reference frame. Let the fluid four-velocity be $u^\mu$ and the local magnetic field be $b^\mu$ (with $b_\mu u^\mu = 0$), then
\begin{equation}
	\chi=\frac{k_\mu(\partial_t)^\mu}{k_\mu u^\mu}=\frac{k_t}{k_\mu u^\mu}=\frac{-1}{k_\mu u^\mu}.
\end{equation}
Here, $k^\mu$ is the four-momentum of the reference photon ($k_{t}=-1$). From the above analysis, it is evident that to calculate the intensity, we must obtain the emissivity and the absorption coefficient.

\subsection{Synchrotron Radiation}
In equation (\ref{eq:rcs}), the radiation coefficients $j_{\nu}$ and $\alpha_{\nu}$ are related to the radiation process, with different radiation processes corresponding to different values of $j_{\nu}$ and $\alpha_{\nu}$. In this paper, we consider synchrotron radiation of electrons under extreme relativistic conditions (in CGS units). In this section, $e$ represents the elementary charge, $c$ is the speed of light, $h$ is the Planck constant, and $k_{B}$ is the Boltzmann constant.

In a plasma system, synchrotron radiation mainly arises from the contribution of electrons, and its emissivity is given by
\begin{equation}
	j_{\nu} = \frac{\sqrt{3} e^{3} b \sin{\theta_{b}}}{4 \pi m_{e} c^{2}} \int_{0}^{\infty} d\gamma \, \mathcal{N}(\gamma) F\left( \frac{\nu}{\nu_{s}} \right),\label{eq:jv}
\end{equation}
where $j_{\nu}$ plays an important role in the imaging of thick disks. Below, we provide a brief introduction to the physical quantities involved. In equation (\ref{eq:jv}), $\gamma = 1/\sqrt{1-\beta^2}$ represents the Lorentz factor of the charged particles, and $\mathcal{N}(\gamma)$ is the distribution function of the electrons. The function $F(x)$ is defined as
\begin{equation}
	F(x) = x \int_x^\infty dy K_{5/3}(y),
\end{equation}
where $K_n(x)$ is the modified Bessel function of the second kind of order $n$. The angle $\theta_{b}$ between $e_{(b)}^{\mu}$ and $e_{(k)}^{\mu}$ is given by
\begin{equation}
	\theta_{b}=\arccos\left(e_{(b)}^{\mu}\cdot e_{(k)}^{\mu}\right)=\arccos\left[\frac{\chi}{b}(b_{\mu}k^{\mu})\right],
\end{equation}
here,
\begin{equation}
	e_{(k)}^{\mu}=-\left(\frac{k^{\mu}}{u^{\nu}k_{\nu}}+u^{\mu}\right),\quad e_{(b)}^{\mu}=\frac{b^{\mu}}{b},
\end{equation}
where $b$ is the magnitude of the local magnetic field. The characteristic frequency $\nu_s$ is given by
\begin{equation}
	\nu_s = \frac{3 e b \sin\theta_b \gamma^2}{4\pi mc}.
\end{equation}
Different electron distributions correspond to different emissivities. In this paper, we consider a thermal distribution, with the distribution function given by
\begin{equation}
	\mathcal{N}(\gamma) = \frac{n_{e} \gamma^{2} \beta}{\Theta_{e} K_{2}(1/\Theta_{e})} e^{(-\gamma/\Theta_{e})},
\end{equation}
here, $n_{e}$ is the electron number density, $\Theta_{e} = k_{B} T_{e} / m_{e} c^{2}$ represents the dimensionless electron temperature, and $T_{e}$ is the electron thermodynamic temperature. For the ultra-relativistic case, i.e., $\beta \approx 1$ and $\Theta_{e} \gg 1$, the asymptotic formula $K_{2}(1/\Theta_{e}) \approx 2 \Theta_{e}^{2}$ holds. Let $z = \gamma / \Theta_{e}$, then
\begin{equation}
	j_{\nu} = \frac{\sqrt{3} n_{e} e^{3} b \sin{\theta_{b}}}{8 \pi m_{e} c^{2}} \int_{0}^{\infty} dz \, z^{2} e^{-z} \, F\left( \frac{\nu}{\nu_{s}} \right).
\end{equation}
Let $x = (\nu / \nu_{s}) z^{2}$, then the emissivity is
\begin{equation}
	j_{\nu} = \frac{n_{e} e^{2} \nu}{2 \sqrt{3} c \Theta_{e}^{2}} \mathcal{F}(x), \quad x = \frac{\nu}{\nu_{c}}, \quad \nu_{c} = \frac{3 e b \sin{\theta_{b}} \Theta_{e}^{2}}{4 \pi m_{e} c},\label{eq:em}
\end{equation}
where the dimensionless function
\begin{equation}
	\mathcal{F}(x) = \frac{1}{x} \int_{0}^{\infty} z^{2} e^{-z} F\left( \frac{x}{ z^{2}} \right),
\end{equation}
cannot be expressed in analytical functions and requires approximation using a fitting function.

This paper will discuss two radiation models: isotropic radiation and anisotropic radiation. For isotropic radiation, we only consider the magnetic field strength while neglecting its direction. The emissivity can be expressed as
\begin{equation}
	j_{\nu} = \frac{1}{2} \int_{0}^{\pi} j_{\nu} \sin\theta_{b}\,\mathrm{d}\theta_{b},
	\label{eq:em2}
\end{equation}
and the corresponding fitting formula is
\begin{equation}
	j_{\nu} = \frac{n e^{2} \nu}{2\sqrt{3}\,c\,\Theta_{e}^{2}}\,\mathcal{F}(x),
	\quad x = \frac{\nu}{\nu_{c}}, \quad 
	\nu_{c} = \frac{3 e b \Theta_{e}^{2}}{4\pi m_{e} c},
\end{equation}
where the fitting function for $\mathcal{F}(x)$ is~\cite{leung2011numerical}
\begin{equation}
	\mathcal{F}(x) =
	\frac{4.0505}{x^{1/6}}
	\left( 1 + \frac{0.4}{x^{1/4}} + \frac{0.5316}{x^{1/2}} \right)
	\exp\!\left( -1.8899\,x^{1/3} \right).
\end{equation}
For anisotropic radiation, we assume that the magnetic field is a mixture of toroidal and poloidal components, which can be expressed as
\begin{equation}
	b^{\mu} \sim (l, 0, 0, 1),
	\label{eq:b}
\end{equation}
where
\begin{equation}
	l = -\frac{u_{\phi}}{u_{t}}, 
	\quad u_{\nu} = g_{\mu\nu} u^{\mu} = (u_{t}, u_{r}, u_{\theta}, u_{\phi}).
\end{equation}
This form of the magnetic field is perpendicular to the fluid four-velocity, i.e., it satisfies $u^{\mu} b_{\mu} = 0$. The emissivity for anisotropic radiation is given by equation~(\ref{eq:em}), and the fitting function for $\mathcal{F}(x)$ is~\cite{mahadevan1996harmony}
\begin{equation}
	\mathcal{F}(x) =
	2.5651 \left( 1 + 1.92\,x^{-1/3} + 0.9977\,x^{-2/3} \right)
	\exp\!\left( -1.8899\,x^{1/3} \right).
\end{equation}

In the context of a thermal electron distribution, the absorption process follows Kirchhoff's law, so the absorption coefficient $\alpha_{\nu}$ satisfies
\begin{equation}
	\alpha_{\nu}=\frac{j_{\nu}}{b_{\nu}},\quad b_{\nu}=\frac{2h\nu^{3}}{c^{2}}\frac{1}{\exp\left(h\nu/k_{B}T_{e}\right)-1},\label{eq:a}
\end{equation}
where $b_{\nu}$ represents the Planck blackbody radiation function.

To simplify calculations, the following constants can be defined in the numerical simulation
\begin{equation}
	\zeta_{1}=\frac{n_{h}e^{2}\nu_{h}}{2\sqrt{3}c\Theta_h^{2}},\quad \zeta_{2}=\frac{4\pi}{3}\frac{m_{e}c\nu_{h}}{eb_{h}\Theta_h^{2}},\quad \zeta_{3}=\frac{h\nu_{h}}{m_{e}c^{2}}\frac{1}{\Theta_h},\quad \zeta_{4}=\frac{2h\nu_{h}^{3}}{c^{2}},\quad \zeta_{5}=(n_{h}m_{p}c^{2})^{1/2}.
\end{equation}
Here, $n_h$ and $\Theta_h$ are evaluated at the event horizon, $\nu_{h} = 10^9 \, \mathrm{Hz} = 1\,\mathrm{GHz}$ and $b_h = 1$. Under this parametrization,
\begin{equation}
	j_{\nu}=\zeta_{1}\tilde{n}_{\mathrm{e}}\tilde{\nu}\left(\tilde{\Theta}_{\mathrm{e}}\right)^{-2}\mathcal{F}(x),\quad x=\frac{\zeta_{2}\tilde{\nu}}{\tilde{b}\sin\theta_{b}\tilde{\Theta}_{\mathrm{e}}^{2}},\quad b_{\nu}=\frac{\zeta_{4}\tilde{\nu}^{3}}{\exp\left(\zeta_{3}\tilde{\nu}/\tilde{\Theta}_{e}\right)-1}.\label{eq:p2}
\end{equation}
where $\tilde{\nu}=\nu/\nu_{h},\tilde{n}_e=n_e/n_h,\tilde{b}=b/b_h,\tilde{\Theta}_e=\Theta_e/\Theta_h$.

Based on equations (\ref{eq:a}) and (\ref{eq:p2}), the intensity in equation (\ref{eq:is}) can, in principle, be calculated, but the electron number density and electron temperature have not yet been determined in the formulas. Below, we will discuss how these parameters are determined in different accretion disk models.

\section{Accretion Disk Models}\label{sec4}
For the non-polarized case, two geometrically thick and optically thin accretion disk models can be considered: the RIAF model~\cite{broderick2011evidence} and the BAAF model~\cite{zhang2024imaging,hou2024new}. The former is highly consistent with general relativistic magnetohydrodynamic (GRMHD) simulations and has successfully reproduced the overall morphological features of M87*~\cite{akiyama2019first}. However, it neglects outflows, non-thermal particles, and the full GRMHD dynamics, thus limiting its application in polarization studies. The latter assumes that the fluid near the event horizon is primarily accelerated by gravity and provides explicit expressions for the thermodynamic variables and magnetic field configuration, offering a better description of the morphology and dynamics of the geometrically thick accretion flow in the near-horizon region of the black hole.

\subsection{RIAF Model}
In cylindrical coordinates, the radius is represented as $R = r\sin\theta$, and the height at the equatorial plane $\theta = \pi/2$ is given by $z = r\cos\theta$. Similar to the construction method of the radiatively inefficient accretion flow (RIAF) model in the literature~\cite{broderick2011evidence}, the density and temperature distribution profiles can be defined as
\begin{equation}
	n_e = n_h\left(\frac {r_h}r\right)^2\exp\left(-\frac{z^2}{2R^2}\right),\quad T_e = T_h\left(\frac {r_h}r\right),
\end{equation}
where $n_h$ and $T_h$ are the electron number density and electron temperature at the outer horizon. The cold magnetization parameter is defined as
\begin{equation}
	\sigma=\frac{b^2}\rho=\frac{b^2}{n_e(m_pc^2)},
\end{equation}
which allows the magnetic field strength to be defined as
\begin{equation}
	b=\sqrt{\sigma\rho},
\end{equation}
where the dimensionless quantity $\rho = n_{e}(m_{p}c^{2})$ represents the fluid mass density. For the accretion disk model considered in this paper, the order of magnitude of $\sigma$ is $\sigma \sim 0.1$~\cite{pu2016effects}. For the RIAF model, both isotropic radiation (\ref{eq:em2}) and anisotropic radiation (\ref{eq:em}) can be considered.

For the accretion flow motion model, this paper considers the infalling motion~\cite{wang2025imaging}. Assuming that the fluid is at rest at infinity, i.e., $u_{t} = -1$, the four-velocity is given by
\begin{equation}
	u^\mu = \left(-g^{tt}, -\sqrt{-(1 + g^{tt})g^{rr}}, 0, -g^{t\phi}\right).\label{eq:im}
\end{equation}
The condition for the four-velocity to be  timelike everywhere requires that $g^{tt} \leq -1$.

\subsection{BAAF Disk Model}
The BAAF model is a steady-state, axisymmetric accretion model~\cite{hou2024new,zhang2024imaging}. In this model,  the fluid is treated as electrically neutral and  the plasma is assumed to be fully ionized into electrons and protons. The accreting particle  is confined to constant $\theta$ surface, i.e., $u^{\theta} \equiv 0$. In this case, the mass conservation equation is
\begin{equation}
	\frac{d}{dr} \left( \sqrt{-g} \rho u^r \right) = 0,
\end{equation}
and its solution is
\begin{equation}
	\rho = \rho_0 \frac{\left.\sqrt{-g}u^r\right|_{r=r_0}}{\sqrt{-g}u^r},
\end{equation}
where $\rho_0 = \rho(r_0)$ is the mass density at the reference point $r_0$. The energy-momentum tensor, projected along $u^\mu$, satisfies
\begin{equation}
	d U = \frac{U + p}{\rho} d \rho ,\label{eq:dedp}
\end{equation}
where $U$ is the internal energy of the fluid. Defining $k = T_p / T_e$ as the temperature ratio between protons and electrons, the internal energy of the fluid in this approximation is
\begin{equation}
	U = \rho + \rho \frac{3}{2} (k+2) \frac{m_e}{m_p} \Theta_e ,\label{eq:U}
\end{equation}
where $\Theta_e = k_B T_e / m_e c^2$ is the dimensionless electron temperature. From the ideal gas equation of state, we have
\begin{equation}
	p = n k_B (T_p + T_e) = \rho (1+k) \frac{m_e}{m_p} \Theta_e .\label{eq:p}
\end{equation}
Substituting equations (\ref{eq:U}) and (\ref{eq:p}) into equation (\ref{eq:dedp}) and integrating, we obtain
\begin{equation}
	\Theta_e = (\Theta_e)_0 \left( \frac{\rho}{\rho_0} \right)^{\frac{2(1+k)}{3(2+k)}} ,
\end{equation}
where $(\Theta_e)_0$ is the reference temperature at $r_h$.

For the convenience of calculation, we assume that $\rho(r_h,\theta)$ follows a Gaussian distribution in the $\theta$ direction, and in the conical solution, we take $\Theta_e(r_h,\theta)$ as a constant
\begin{equation}
	\rho(r_h,\theta) = \rho_h \exp \left[ - \left( \frac{\sin \theta - \sin \mu_\theta}{\sigma_\theta} \right)^2 \right], 
	\quad 
	\Theta(r_h,\theta) = \Theta_h ,\label{eq:th}
\end{equation}
where $\mu_\theta$ represents the mean position in the $\theta$ direction, and $\sigma_\theta$ is the standard deviation of the distribution. For M87*, observational results indicate that $\rho_h \approx 1.5 \times 10^{3} \,\mathrm{g/cm/s^{2}}$ and $\Theta_h \approx 16.86$, corresponding to $n_h = 10^6 \,\mathrm{cm^{-3}}$ and $T_h = 10^{11} \,\mathrm{K}$~\cite{vincent2022images}.

In a steady-state axisymmetric spacetime, the general form of the magnetic field is given by
\begin{equation}
	b^{\mu} = \frac{\Phi}{\sqrt{-g}u^{r}}\left[\left(u_{t} + \Omega_{b} u_{\phi}\right)u^{\mu} + \delta_{t}^{\mu} + \Omega_{b} \delta_{\phi}^{\mu}\right],\label{eq:b2}
\end{equation}
where $\Phi = F_{\theta\phi}$ is a component of the electromagnetic field tensor, and can be expressed as
\begin{equation}
	\Phi = \Phi_0 \,\mathrm{sign}(\cos \theta)\,\sin \theta.
\end{equation}
$\Omega_{b}$ is the angular velocity of the magnetic field, and in this paper, we take $\Omega_{b} = 0.3 \Omega_{h}$, where $\Omega_{h} ={a}/(2r_h)$ is the angular velocity of the rotating black hole. For the BAAF model, only anisotropic radiation (\ref{eq:em}) is considered. For the fluid's four-velocity, we can use the ballistic approximation, where the fluid moves along the geodesics \cite{hou2024new}, or the infalling motion  mentioned in (\ref{eq:im}). We will choose the later in this paper. 

\section{Polarization Imaging}\label{sec5}
In the WKB approximation, the propagation of light satisfies the radiative transfer equation
\begin{equation}
	k^\mu \nabla_\mu \hat{S}^{\alpha\beta} = \hat{J}^{\alpha\beta} + \hat{H}^{\alpha\beta\mu\nu} \hat{S}_{\mu\nu},\label{eq:rte2}
\end{equation}
where $k^\mu$ and $\hat{S}^{\alpha\beta}$ are the photon wave vector and the polarization tensor, respectively, and $\hat{J}^{\alpha\beta}$ and $\hat{H}^{\alpha\beta\mu\nu}$ describe the emission properties of the radiation source and the effects of the medium on photon propagation, the latter often including absorption and Faraday rotation effects~\cite{huang2024coport}. By using the gauge invariance of $\hat{S}^{\alpha\beta}$, calculations can be carried out in a simple parallel transport frame. The covariant form of the radiative transfer equation (\ref{eq:rte2}) is then decomposed into two parts. The first part reflects the gravitational effects
\begin{equation}
	k^\mu \nabla_\mu f^\nu = 0, \quad f^\mu k_\mu = 0 ,
\end{equation}
where $f^\mu$ is a normalized spacelike vector perpendicular to $k^\mu$. The second part is the radiative transfer equation
\begin{equation}
	\frac{d}{d\lambda} \hat{S} = \hat{R}(\vartheta)\hat{J} - \hat{R}(\vartheta)\hat{M}\hat{R}(-\vartheta)\hat{S} ,
\end{equation}
where
\begin{equation}
	\hat{S} = \begin{pmatrix} \tilde{I} \\ \tilde{Q} \\ \tilde{U} \\ \tilde{V} \end{pmatrix}, \quad \hat{J} = \frac{1}{\nu^2} \begin{pmatrix} j_I \\ j_Q \\ j_U \\ j_V \end{pmatrix}, \quad \hat{M} = \nu \begin{pmatrix} a_I & a_Q & a_U & a_V \\ a_Q & a_I & r_V & -r_U \\ a_U & -r_V & a_I & r_Q \\ a_V & r_U & -r_Q & a_I \end{pmatrix}.
\end{equation}
$\hat{R}(\vartheta)$ is the rotation matrix
\begin{equation}
	\hat{R}(\vartheta) =
	\begin{pmatrix}
		1 & 0 & 0 & 0 \\
		0 & \cos(2\vartheta) & -\sin(2\vartheta) & 0 \\
		0 & \sin(2\vartheta) & \cos(2\vartheta) & 0 \\
		0 & 0 & 0 & 1
	\end{pmatrix}, \label{eq:rx}
\end{equation}
where the rotation angle $\vartheta$ is the angle between the reference vector $f^\mu$ and the local magnetic field $b^\mu$ in the transverse subspace of the photon trajectory, given by
\begin{equation}
	\vartheta = \mathrm{sign}(\epsilon_{\alpha\beta\mu\nu} u^\alpha f^\beta b^\mu k^\nu)
	\arccos \left( \frac{h^{\mu\nu} f_\mu b_\nu}{\sqrt{(h^{\mu\nu} f_\mu f_\nu)(h^{\alpha\beta} b_\alpha b_\beta)}} \right) ,
\end{equation}
where $h^{\mu\nu}$ is the induced metric in the transverse subspace. At the observer’s location, the Stokes parameters need to be projected onto the observer’s screen, which can be done using the rotation matrix (\ref{eq:rx}), with the rotation angle being
\begin{equation}
	\vartheta_o = \mathrm{sign}(\epsilon_{\alpha\beta\mu\nu} u^{\alpha} f^{\beta} d^{\mu} k^{\nu}) \arccos \left( \frac{h^{\mu\nu} f_\mu d_\nu}{\sqrt{(h^{\mu\nu} f_\mu f_\nu)(h^{\alpha\beta} d_\alpha d_\beta)}} \right),
\end{equation}
where $d^\mu$ is the $y$-axis direction of the screen, and we choose $d^\mu = -\partial_{\theta}^\mu$. The projection results in four Stokes parameters
\begin{equation}
	\tilde{I}_o = \tilde{I}, \quad \tilde{Q}_o = \tilde{Q} \cos \vartheta_o - \tilde{U} \sin \vartheta_o, \quad \tilde{U}_o = \tilde{Q} \sin \vartheta_o + \tilde{U} \cos \vartheta_o, \quad \tilde{V}_o = \tilde{V}.
\end{equation}
Here, $\tilde{I}_o$ reflects the intensity of the light, $\tilde{Q}_o$ and $\tilde{U}_o$ are related to the electric field $\vec{E} = (E_x, E_y)$ as
\begin{equation}
	\tilde{Q}_o = E_x^2 - E_y^2, \quad \tilde{U}_o = 2 E_x E_y.
\end{equation}
If $\tilde{U}_o$ is positive, then $E_x$ and $E_y$ have the same sign, and $\vec{E}$ lies in the first and third quadrants; if $\tilde{U}_o$ is negative, then $E_x$ and $E_y$ have opposite signs, and $\vec{E}$ lies in the second and fourth quadrants. The sign of $\tilde{Q}_o$ reflects the relationship between $\vec{E}$ and the lines $y = x$ or $y = -x$. $\tilde{V}_o$ reflects information about circular polarization: if $\tilde{V}_o$ is positive, it corresponds to left-handed circular polarization, and if negative, to right-handed circular polarization. Through the Stokes parameters, the magnitude and direction of the projected polarization vector $\vec{f}$ on the screen can be determined. The magnitude corresponds to the degree of linear polarization, and the angle corresponds to the electric vector position angle (EVPA)
\begin{equation}
	|\vec{f}| = \mathcal{P}_o = \frac{\sqrt{\tilde{Q}_o^2 + \tilde{U}_o^2}}{\tilde{I}_o}, 
	\qquad 
	\arg(\vec{f}) = \Phi_{\mathrm{EVPA}} 
	= \frac{1}{2} \arctan \left( \frac{\tilde{U}_o}{\tilde{Q}_o} \right).
	\label{eq:pv}
\end{equation}
Through the above analysis, the Stokes parameters and $\vec{f}$ can be computed to obtain the complete polarization characteristics.

\section{Numerical Results}\label{sec6}
Through the ray-tracing method, combined with the ZAMO frame and celestial coordinates, the relationship between the pixel coordinates on the projection screen and the celestial coordinates can be effectively established. This process provides a solid computational foundation for research such as black hole imaging. The specific steps and details can be found in the related literature~\cite{hou2022image,zhong2021qed,wang2025semi,hu2021qed,he2026shadow}.  Throughout the paper, the observation frequency is fixed at $230\,\mathrm{GHz}$, and the accretion flow motion model is the infalling motion. 

\subsection{RIAF Model under Isotropic Radiation}
\begin{figure}[!h]
	\centering 
	\subfigure[$a=0.001$]{\includegraphics[scale=0.35]{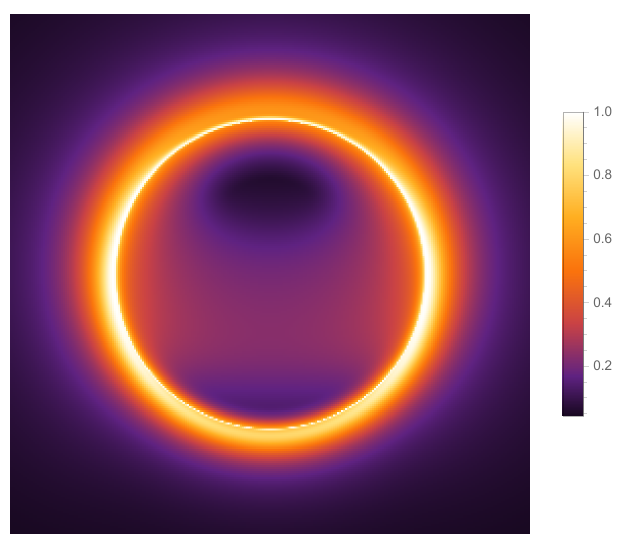}}
	\subfigure[$a=0.1$]{\includegraphics[scale=0.35]{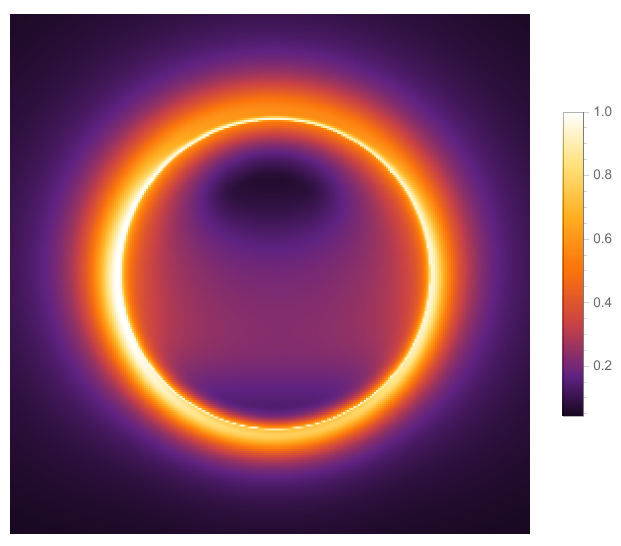}}
	\subfigure[$a=0.5$]{\includegraphics[scale=0.35]{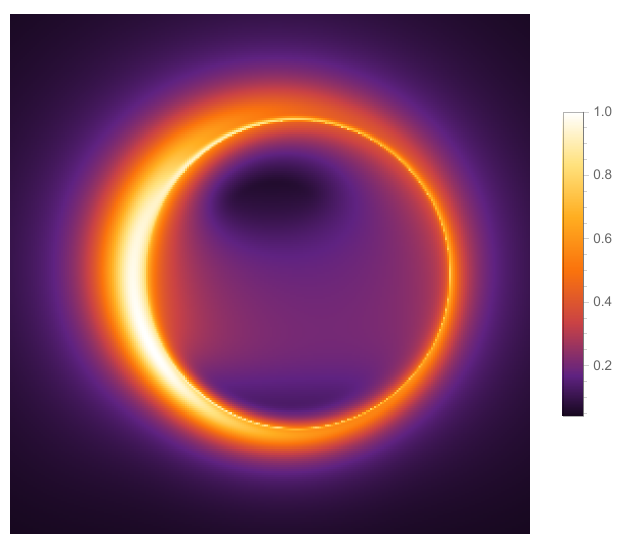}}
	\subfigure[$a=0.98$]{\includegraphics[scale=0.35]{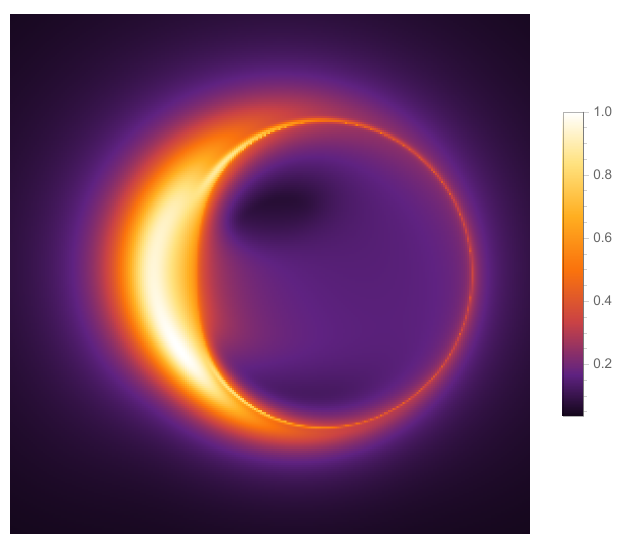}}
	
	\caption{The impact of the spin parameter $a$ on the RIAF model under isotropic radiation. Fixed parameters are  $\varpi=\varsigma=0.1$, and $\theta_{o}=65^\circ$.}
	\label{fig1}
\end{figure}

We first discuss the impact of the spin parameter $a$ on the RIAF model under isotropic radiation. We set the spontaneous Lorentz violating parameters $\varpi = \varsigma = 0.1$, the observer inclination angle is  $\theta_o = 65^\circ$. The numerical results are shown in Figure~\ref{fig1}. We observe a bright ring structure in all the images, corresponding to higher-order images, where photons orbit the black hole one or more times before reaching the observer. These higher-order images are a result of the strong gravitational lensing effect, where the black hole's gravity bends the light path, allowing some photons to bypass the black hole and reach the observer. Beyond this ring structure, the region with nonzero intensity corresponds to the primary image, where photons travel directly from the accretion disk to the observer. Notably, for all parameters, there is a region of decreasing intensity inside the higher-order images, which originates from the black hole’s event horizon. In the case of a geometrically thin accretion disk, the particle  is confined to the equatorial plane, so the event horizon appears as a clearly defined black region in the image (the "inner shadow"), which may be captured by the EHT~\cite{chael2021observing}. However, for a geometrically thick accretion disk, this region may be obscured by radiation from outside the equatorial plane, making it difficult to distinguish. Compared to thin disks, thick disks are more physically realistic, suggesting that direct imaging of the black hole’s event horizon remains challenging.

\begin{figure}[!h]
	\centering 
	\subfigure[$a=0.001$]{\includegraphics[scale=0.35]{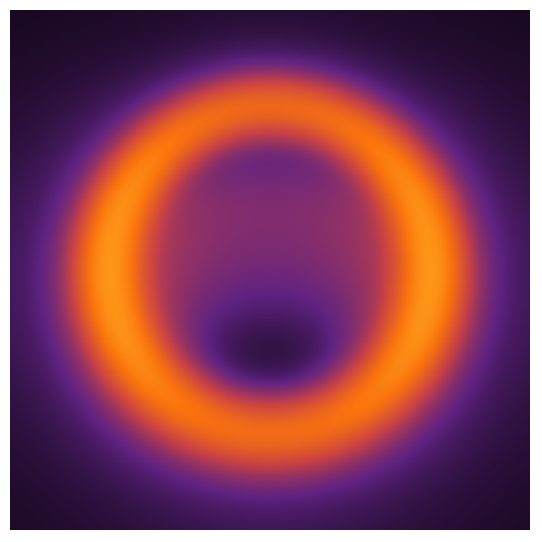}}
	\hspace{0.35cm} 
	\subfigure[$a=0.1$]{\includegraphics[scale=0.35]{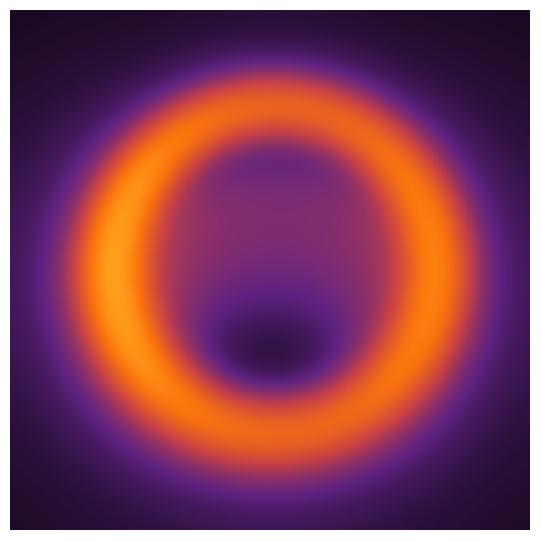}}
	\hspace{0.35cm}
	\subfigure[$a=0.5$]{\includegraphics[scale=0.35]{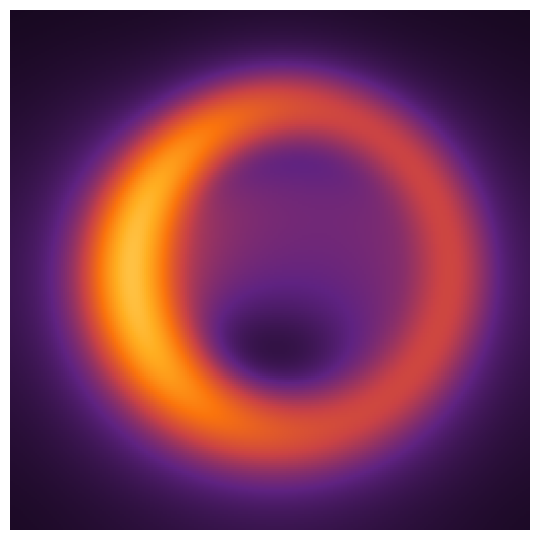}}
	\hspace{0.35cm}
	\subfigure[$a=0.98$]{\includegraphics[scale=0.35]{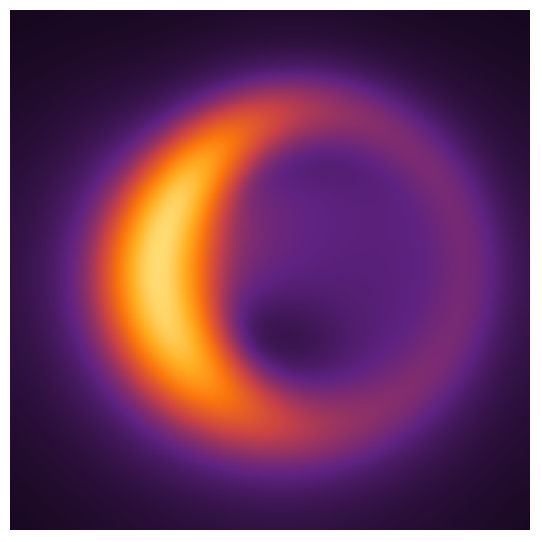}}
	
	\caption{Blurred images of black holes processed with a Gaussian filter, where the standard deviation is set to $1/12$ of the field of view $\gamma_{\mathrm{fov}}$. The plotting parameters are consistent with those in Figure~\ref{fig1}.}
	\label{fig2}
\end{figure}

In Figure~\ref{fig1}, as the spin parameter $a$ increases, the higher-order images gradually transform into a "D"-shaped structure, with the intensity on the left side increasing. When $a = 0.98$, a crescent-shaped bright region appears on the left. This phenomenon originates from the coordinate frame dragging effect; as $a$ increases, the frame dragging effect becomes stronger, and the asymmetry of the higher-order images increases. To match the current EHT's limited resolution, we show the blurred images in Figure~\ref{fig2}, with the same parameters as in Figure~\ref{fig1}. The black hole blurred images are obtained through Gaussian filtering, where the standard deviation of the Gaussian function is $1/12$ of the field of view $\gamma_{\mathrm{fov}}$. In Figure~\ref{fig2}, the boundary between the higher-order images and the primary image becomes blurred, and the distinguishability of the event horizon’s outline is reduced. This further illustrates that, as one of the most iconic structures of a black hole, the event horizon remains challenging to directly verify from an observational standpoint.

\begin{figure}[!h]
	\centering 
	\subfigure[$\varpi=0.3,\varsigma=0.69$]{\includegraphics[scale=0.35]{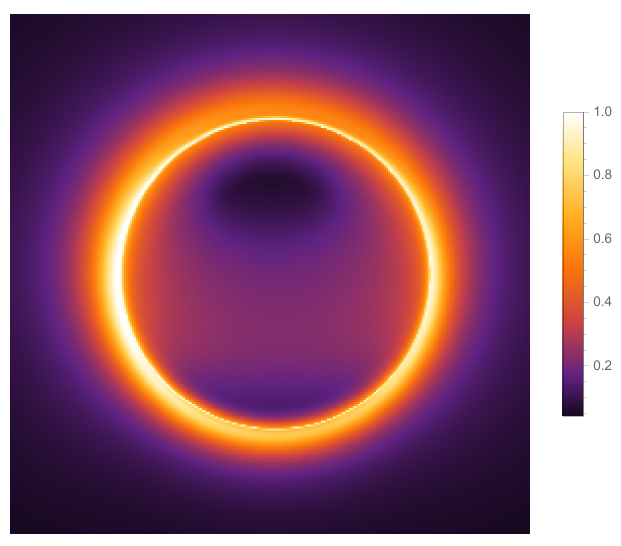}}
	\subfigure[$\varpi=0.3,\varsigma=0.79$]{\includegraphics[scale=0.35]{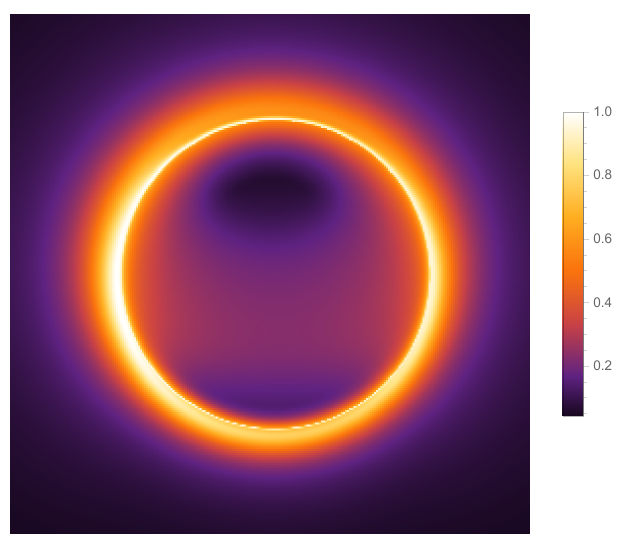}}
	\subfigure[$\varpi=0.3,\varsigma=0.89$]{\includegraphics[scale=0.35]{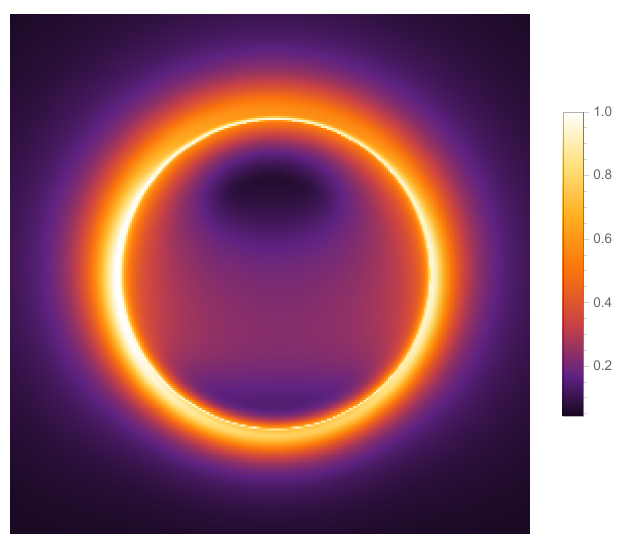}}
	\subfigure[$\varpi=0.3,\varsigma=0.99$]{\includegraphics[scale=0.35]{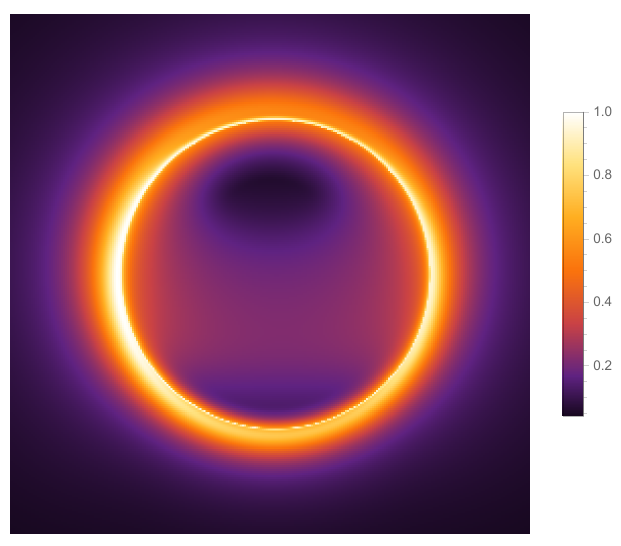}}
	
	\subfigure[$\varpi=0.6,\varsigma=0.69$]{\includegraphics[scale=0.35]{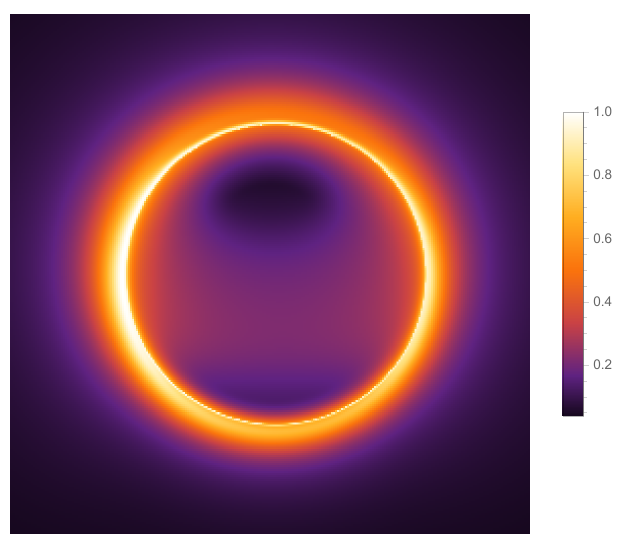}}
	\subfigure[$\varpi=0.6,\varsigma=0.79$]{\includegraphics[scale=0.35]{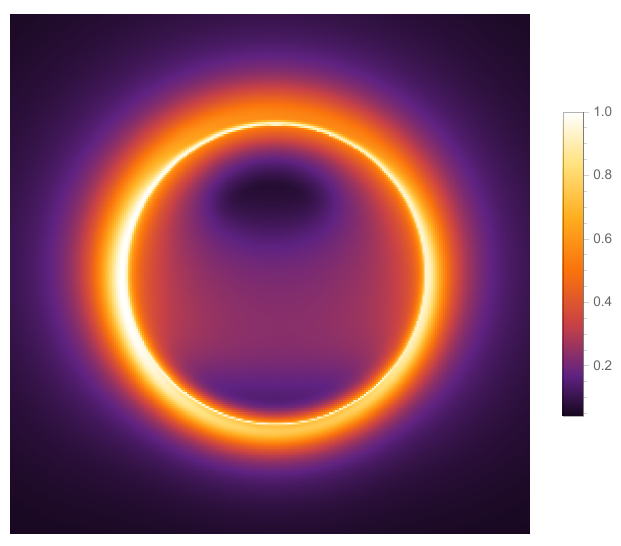}}
	\subfigure[$\varpi=0.6,\varsigma=0.89$]{\includegraphics[scale=0.35]{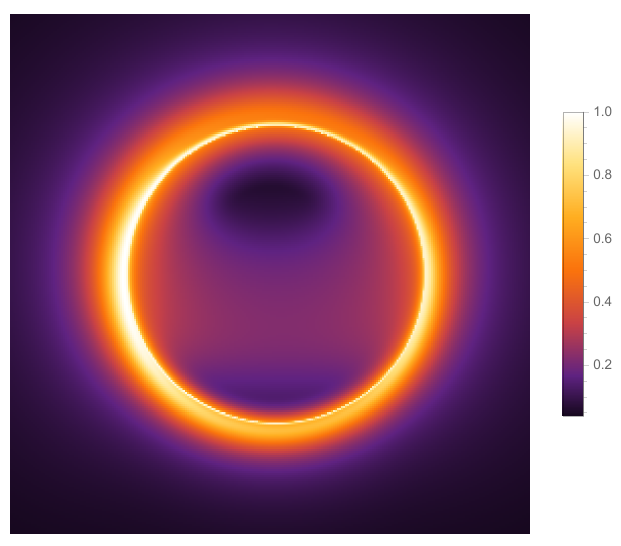}}
	\subfigure[$\varpi=0.6,\varsigma=0.99$]{\includegraphics[scale=0.35]{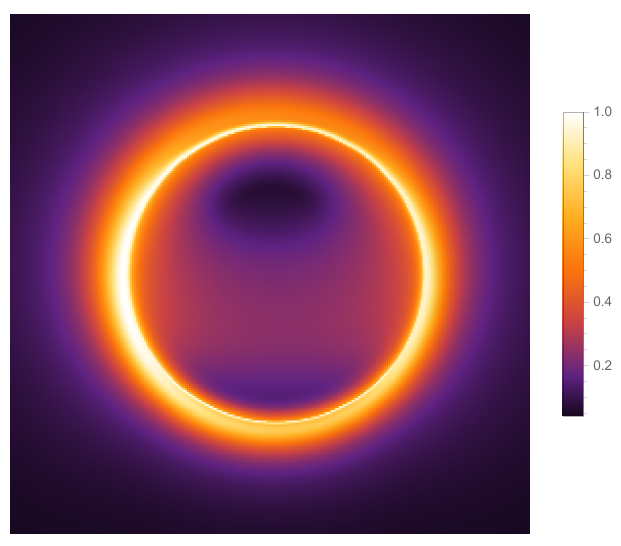}}
	
	\subfigure[$\varpi=0.9,\varsigma=0.69$]{\includegraphics[scale=0.35]{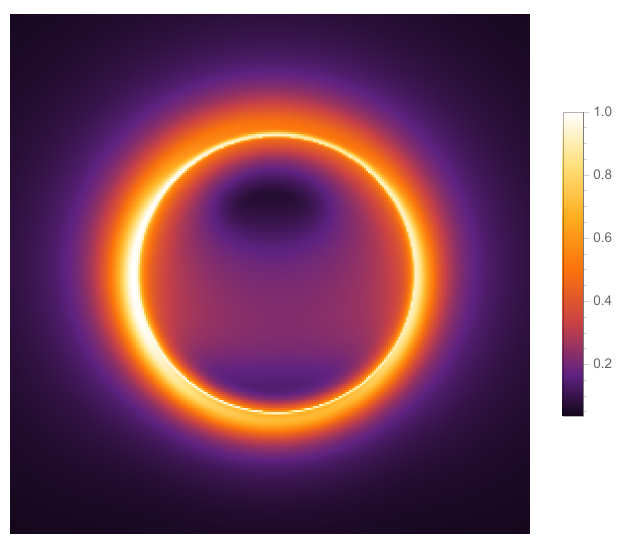}}
	\subfigure[$\varpi=0.9,\varsigma=0.79$]{\includegraphics[scale=0.35]{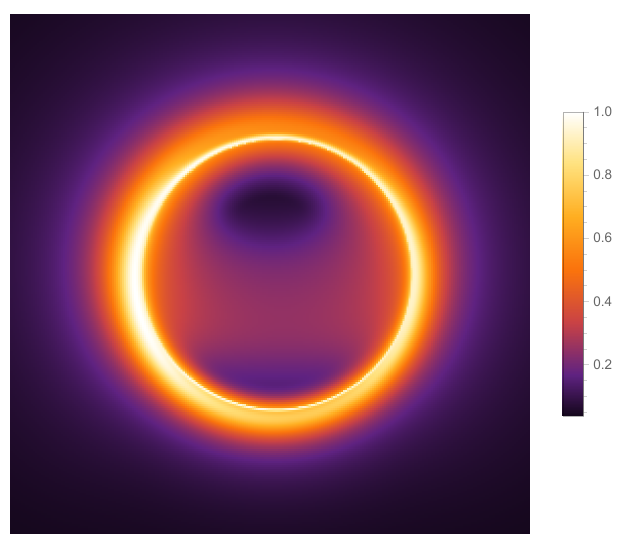}}
	\subfigure[$\varpi=0.9,\varsigma=0.89$]{\includegraphics[scale=0.35]{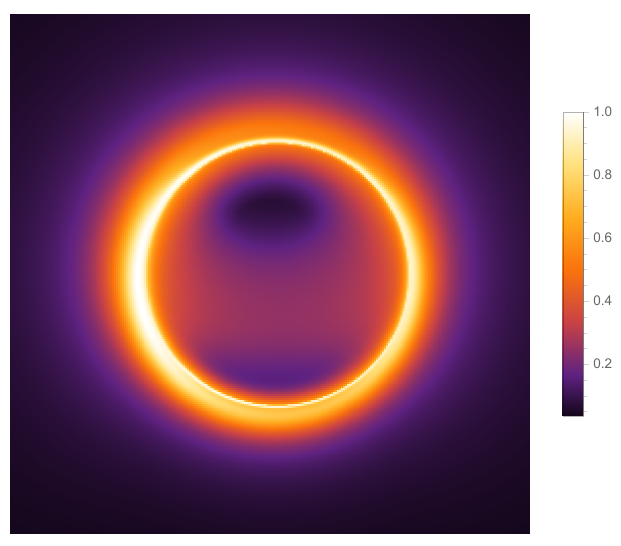}}
	\subfigure[$\varpi=0.9,\varsigma=0.99$]{\includegraphics[scale=0.35]{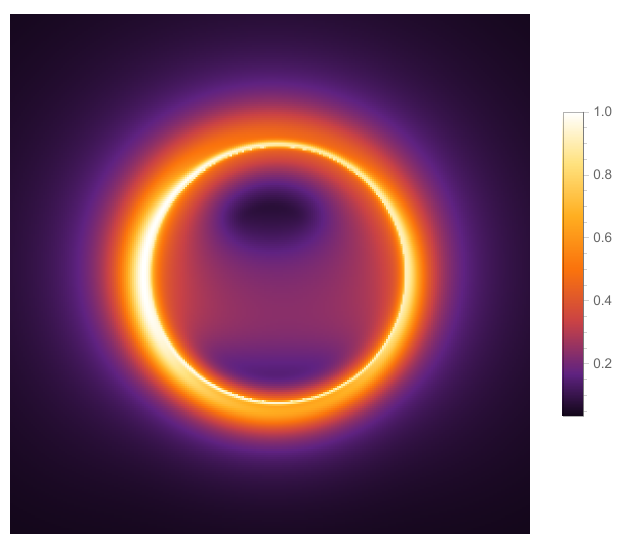}}
	
	\caption{Black hole shadow images under the RIAF model with isotropic radiation. The parameters are fixed at $a = 0.1$, $\theta_o = 65^\circ$.}
	\label{fig3}
\end{figure}

\begin{figure}[!h]
	\centering 
	\subfigure[the x-axis]{\includegraphics[scale=0.8]{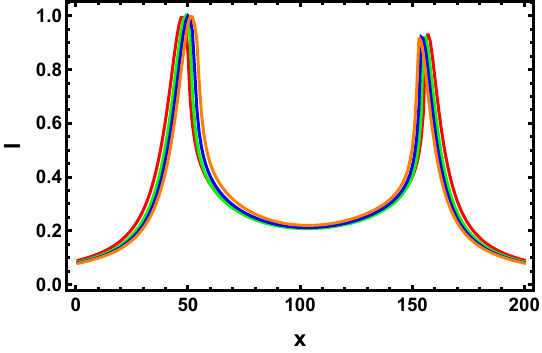}}
	\subfigure[the y-axis]{\includegraphics[scale=0.8]{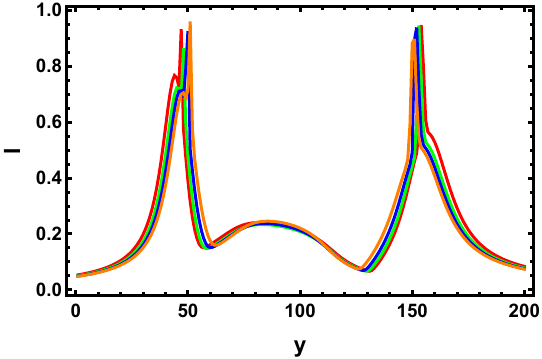}}
	
	\caption{Intensity distributions along the x-axis and y-axis under the RIAF model with isotropic radiation. The red, green, blue, and orange colors correspond to $\varsigma = 0.69, 0.79, 0.89, 0.99$, respectively. The parameters are fixed at $a = 0.1$, $\varpi = 0.9$, $\theta = 65^\circ$.}
	\label{fig4}
\end{figure}

\begin{figure}[!h]
	\centering 
	\subfigure[the x-axis]{\includegraphics[scale=0.8]{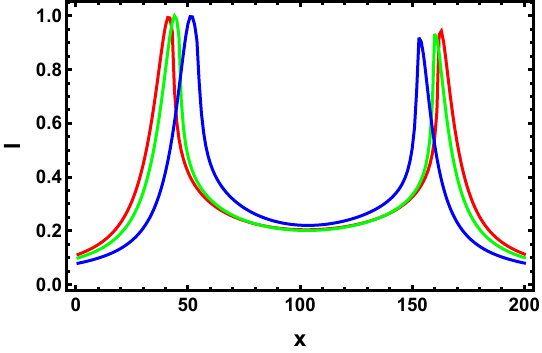}}
	\subfigure[the y-axis]{\includegraphics[scale=0.8]{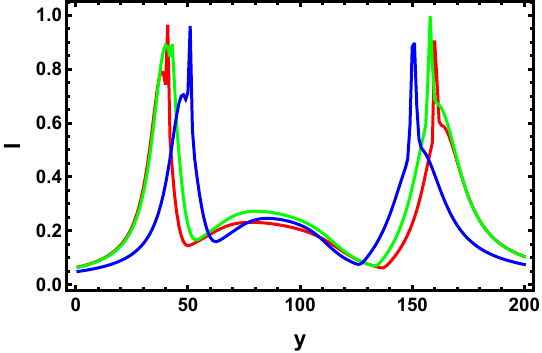}}
	
	\caption{Intensity distributions along the x-axis and y-axis under the RIAF model with isotropic radiation. The red, green, and blue colors correspond to $\varpi = 0.3, 0.6, 0.9$, respectively. The parameters are fixed at $a = 0.1$, $\varsigma = 0.99$, $\theta_o = 65^\circ$.}
	\label{fig5}
\end{figure}

Figure~\ref{fig3} shows the impact of the spontaneous Lorentz violating parameters $\varpi$ and $\varsigma$ on the shadow image, with fixed parameters $a = 0.1$, $\theta_o = 65^\circ$. The images reveal that increasing $\varpi$ and $\varsigma$ has almost no effect on the shape of the higher-order images but reduces their size. Due to the coordinate frame dragging effect, the intensity on the left side of the higher-order images is slightly greater than on the right side. In all images, two dark regions appear within the higher-order images, with the upper region being slightly darker than the lower one. This phenomenon originates from the gravitational lensing effect. To illustrate the effect of $\varpi$ and $\varsigma$ on the image more clearly, we have plotted the intensity distributions along $y = 0$ (the x-axis) and $x = 0$ (the y-axis) in Figures~\ref{fig4} and~\ref{fig5}. In all the intensity cut diagrams, each curve exhibits two prominent peaks corresponding to the higher-order images, while the regions beyond the peaks correspond to the primary image. As $\varpi$ and $\varsigma$ increase, the peaks move closer together, indicating a decrease in the size of the higher-order images. For the intensity distribution along the x-axis, no region with zero intensity is observed, which is due to the radiation from outside the equatorial plane and the observer inclination angle $\theta_o = 65^\circ$. For the intensity distribution along the y-axis, two local minima appear between the two peaks, corresponding to the event horizon. A local maximum appears between the two minima, which is also due to the influence of radiation from outside the equatorial plane.

\subsection{RIAF Model under Anisotropic Radiation}
\begin{figure}[!h]
	\centering 
	\subfigure[$a=0.001$]{\includegraphics[scale=0.35]{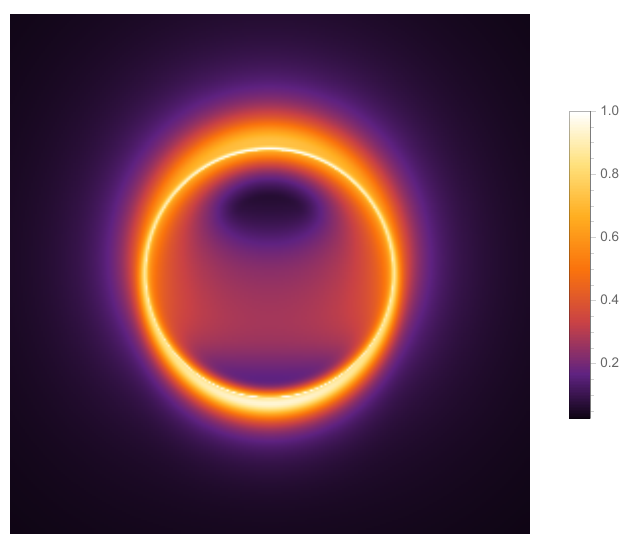}}
	\subfigure[$a=0.1$]{\includegraphics[scale=0.35]{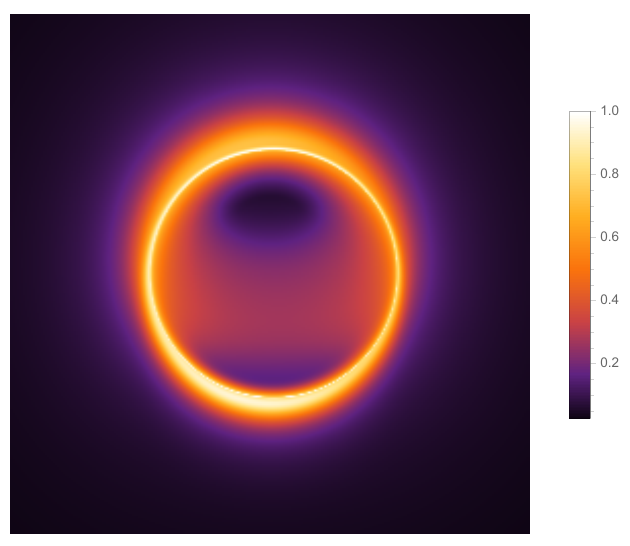}}
	\subfigure[$a=0.5$]{\includegraphics[scale=0.35]{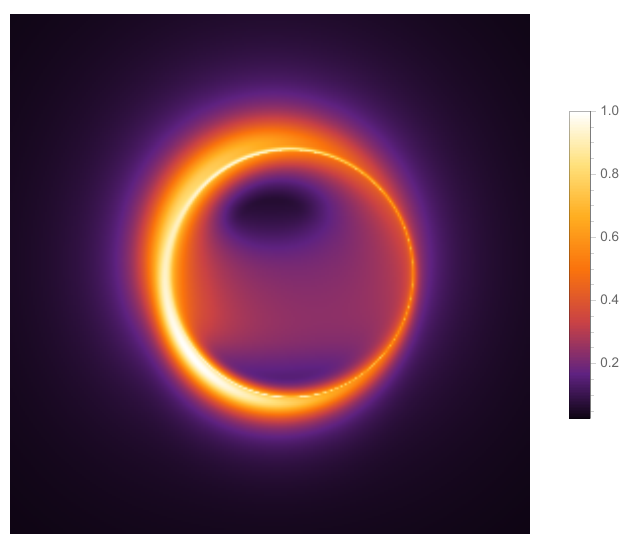}}
	\subfigure[$a=0.98$]{\includegraphics[scale=0.35]{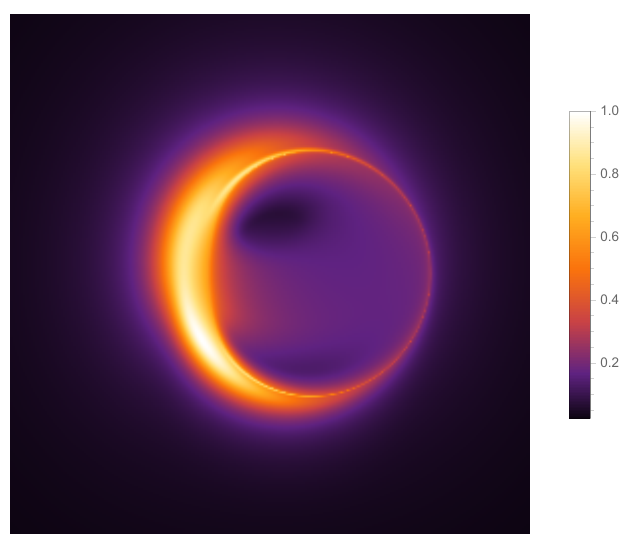}}
	
	\caption{The impact of the spin parameter $a$ on the RIAF model under anisotropic radiation. The fixed parameters are  $\varpi = \varsigma = 0.1$, and $\theta_o = 65^\circ$.}
	\label{fig8}
\end{figure}

\begin{figure}[!h]
	\centering 
	\subfigure[$a=0.001$]{\includegraphics[scale=0.35]{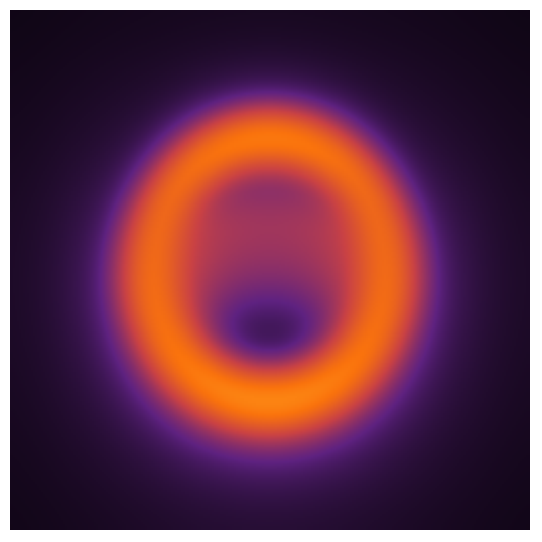}}
	\hspace{0.35cm} 
	\subfigure[$a=0.1$]{\includegraphics[scale=0.35]{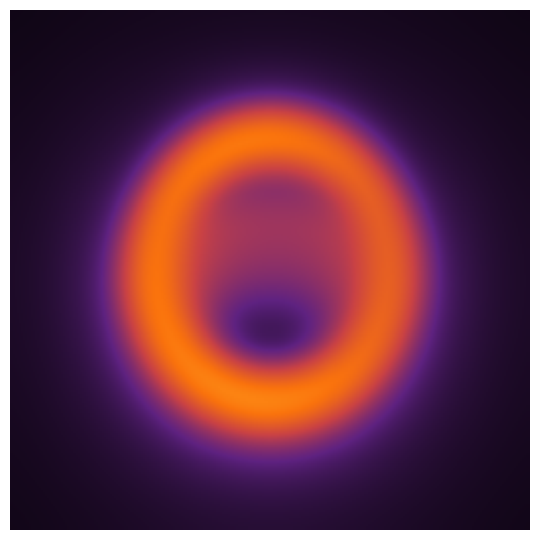}}
	\hspace{0.35cm} 
	\subfigure[$a=0.5$]{\includegraphics[scale=0.35]{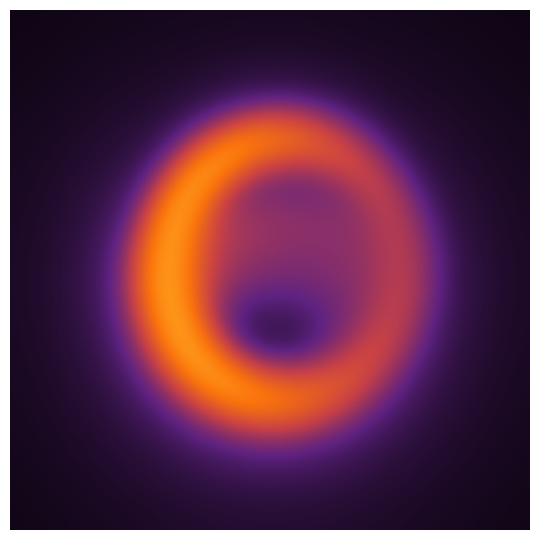}}
	\hspace{0.35cm} 
	\subfigure[$a=0.98$]{\includegraphics[scale=0.35]{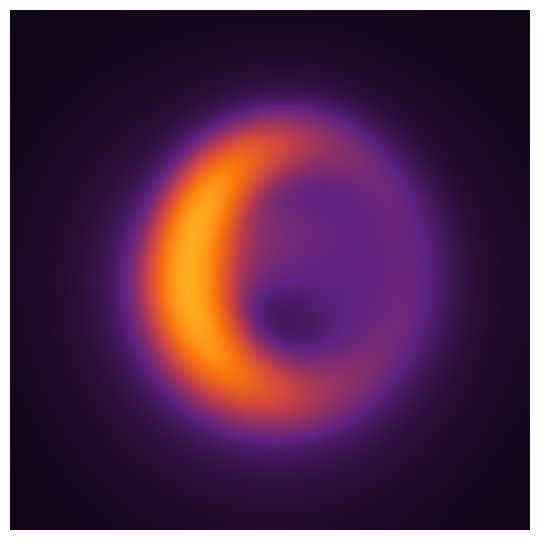}}
	
	\caption{Blurred images of black holes processed with a Gaussian filter, where the standard deviation is set to $1/12$ of the field of view $\gamma_{\mathrm{fov}}$. The plotting parameters are consistent with those in Figure~\ref{fig8}.}
	\label{fig9}
\end{figure}

\begin{figure}[!h]
	\centering 
	\subfigure[$\varpi=0.3,\varsigma=0.69$]{\includegraphics[scale=0.35]{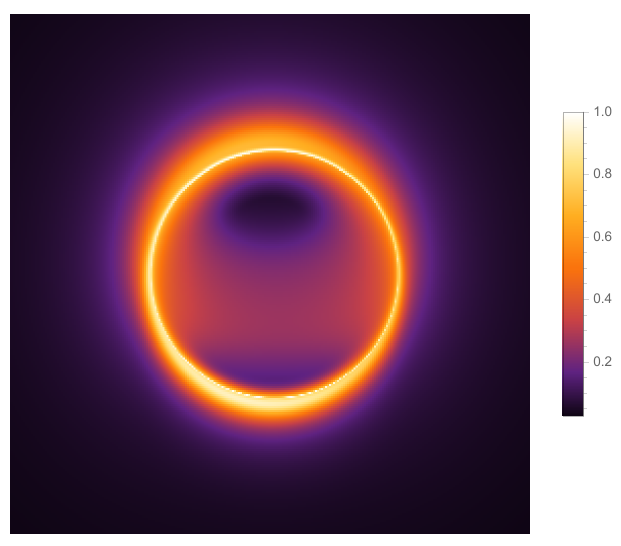}}
	\subfigure[$\varpi=0.3,\varsigma=0.79$]{\includegraphics[scale=0.35]{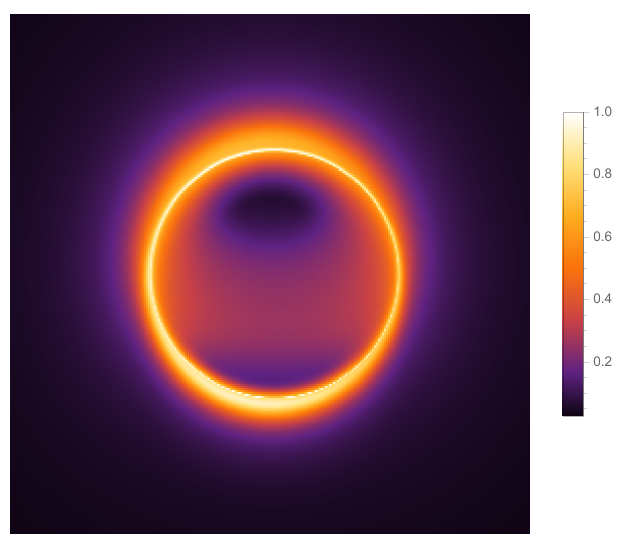}}
	\subfigure[$\varpi=0.3,\varsigma=0.89$]{\includegraphics[scale=0.35]{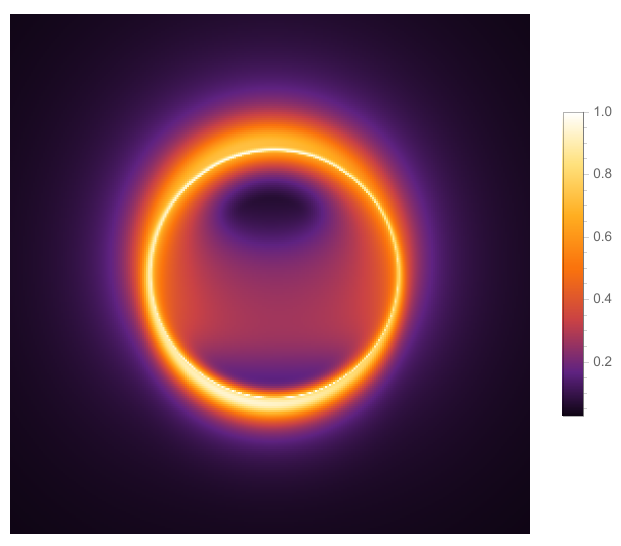}}
	\subfigure[$\varpi=0.3,\varsigma=0.99$]{\includegraphics[scale=0.35]{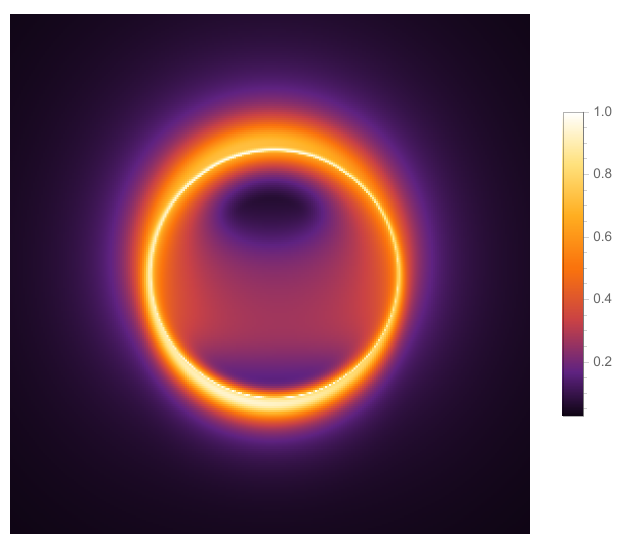}}
	
	\subfigure[$\varpi=0.6,\varsigma=0.69$]{\includegraphics[scale=0.35]{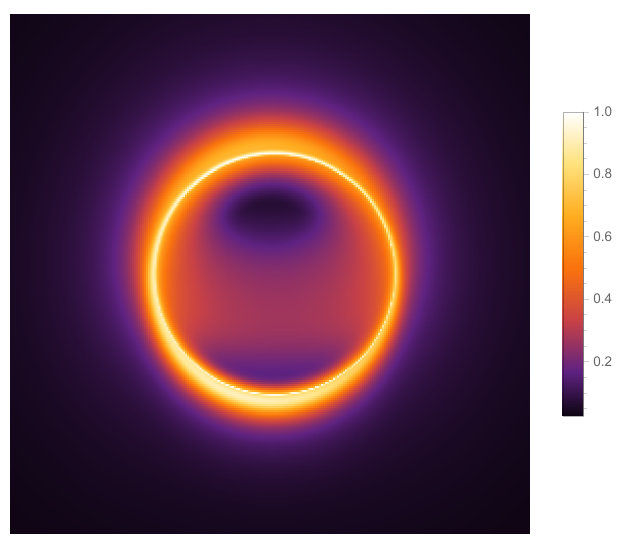}}
	\subfigure[$\varpi=0.6,\varsigma=0.79$]{\includegraphics[scale=0.35]{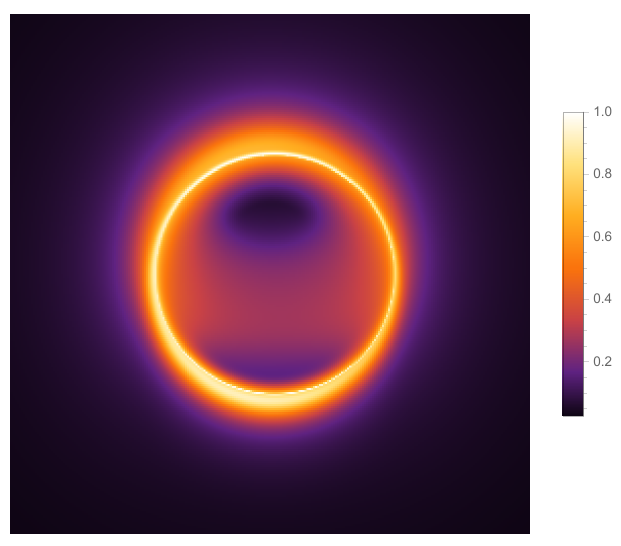}}
	\subfigure[$\varpi=0.6,\varsigma=0.89$]{\includegraphics[scale=0.35]{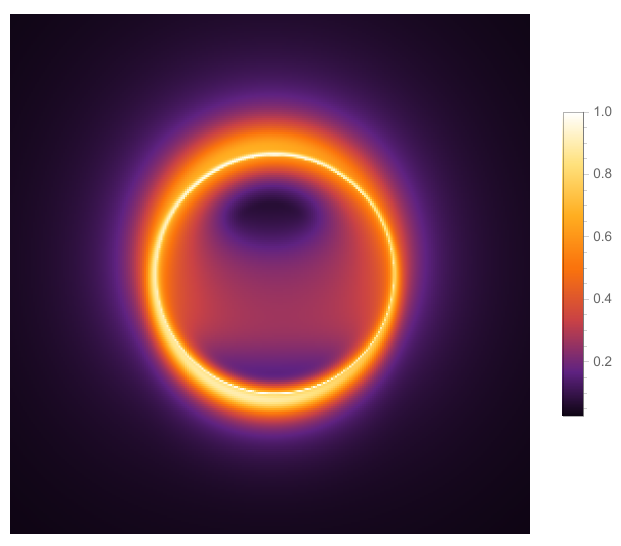}}
	\subfigure[$\varpi=0.6,\varsigma=0.99$]{\includegraphics[scale=0.35]{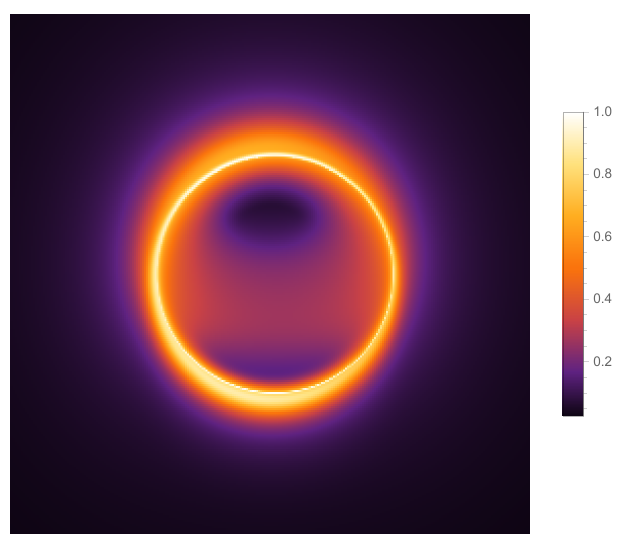}}
	
	\subfigure[$\varpi=0.9,\varsigma=0.69$]{\includegraphics[scale=0.35]{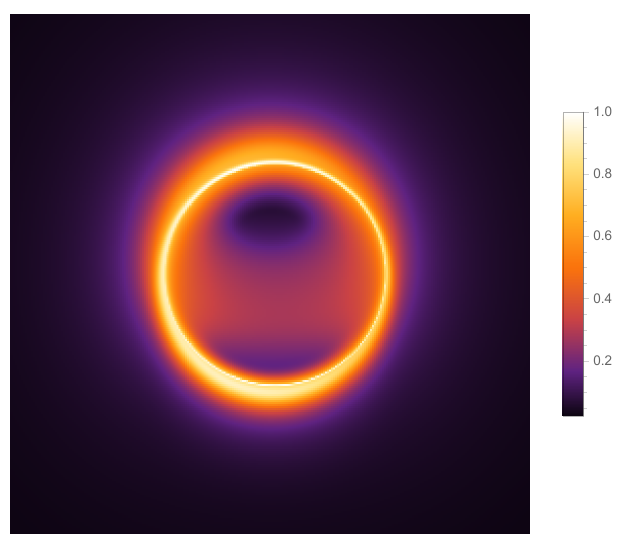}}
	\subfigure[$\varpi=0.9,\varsigma=0.79$]{\includegraphics[scale=0.35]{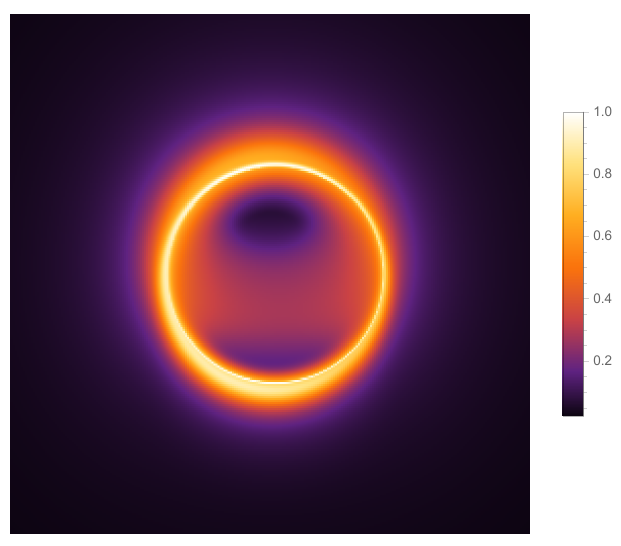}}
	\subfigure[$\varpi=0.9,\varsigma=0.89$]{\includegraphics[scale=0.35]{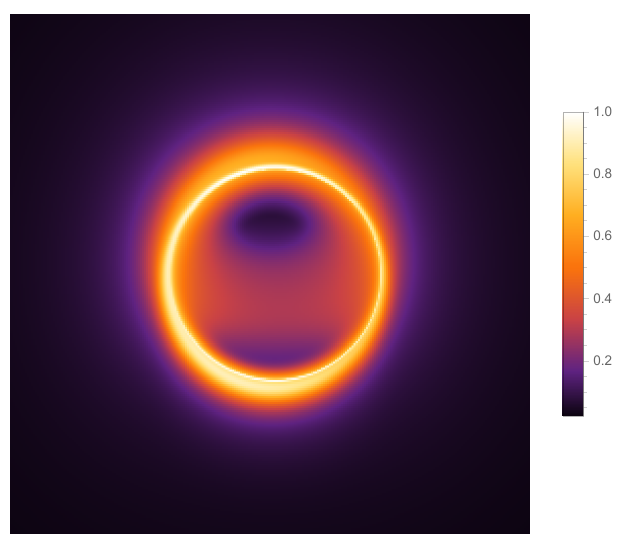}}
	\subfigure[$\varpi=0.9,\varsigma=0.99$]{\includegraphics[scale=0.35]{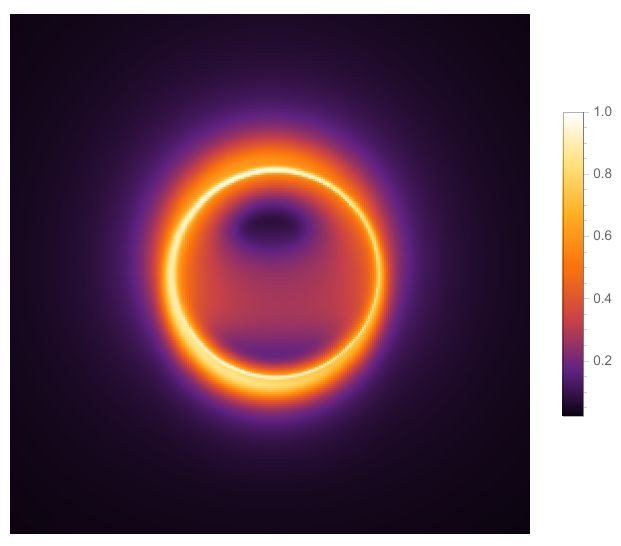}}
	
	\caption{Black hole shadow images under the RIAF model with anisotropic radiation. The fixed parameters are $a = 0.1$, $\theta_o = 65^\circ$.}
	\label{fig10}
\end{figure}

\begin{figure}[!h]
	\centering 
	\subfigure[the x-axis]{\includegraphics[scale=0.8]{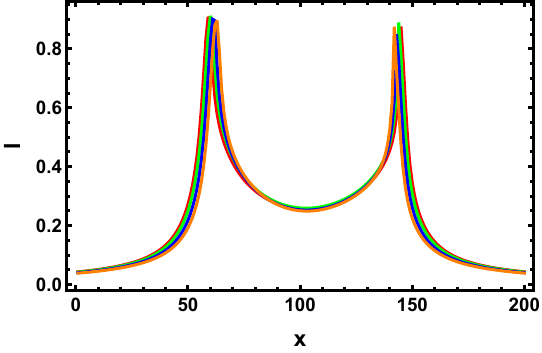}}
	\subfigure[the y-axis]{\includegraphics[scale=0.8]{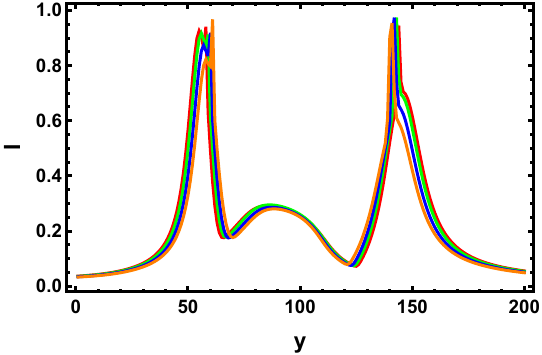}}
	
	\caption{Intensity distributions along the x-axis and y-axis under the RIAF model with anisotropic radiation. The red, green, blue, and orange colors correspond to $\varsigma = 0.69, 0.79, 0.89, 0.99$, respectively. The fixed parameters are $a = 0.1$, $\varpi = 0.9$, $\theta = 65^\circ$.}
	\label{fig11}
\end{figure}

\begin{figure}[!h]
	\centering 
	\subfigure[the x-axis]{\includegraphics[scale=0.8]{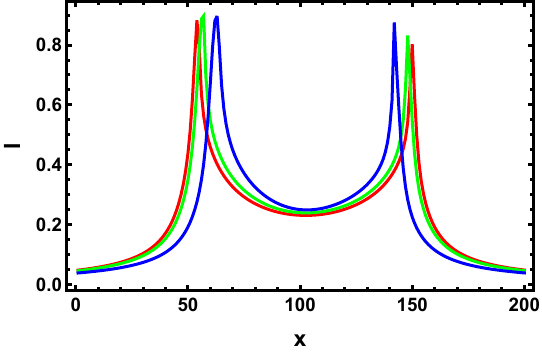}}
	\subfigure[the y-axis]{\includegraphics[scale=0.8]{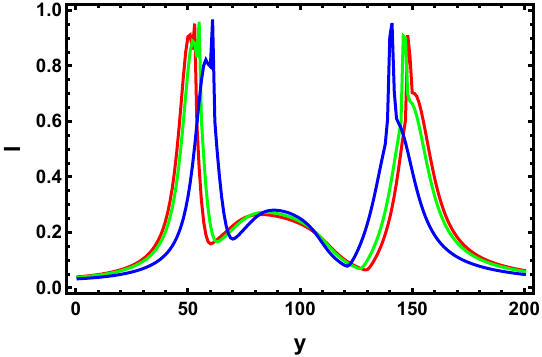}}
	
	\caption{Intensity distributions along the x-axis and y-axis under the RIAF model with anisotropic radiation. The red, green, and blue colors correspond to $\varpi = 0.3, 0.6, 0.9$, respectively. The fixed parameters are $a = 0.1$, $\varsigma = 0.99$, $\theta_o = 65^\circ$.}
	\label{fig12}
\end{figure}

Figures~\ref{fig8} and~\ref{fig9} show the impact of the spin parameter $a$ on the RIAF model under anisotropic radiation, along with the corresponding blurred images. For isotropic radiation, when $a$ is small, the higher-order images are approximately circular, but under anisotropic radiation, they significantly deform into elliptical shapes. The main reason for this difference is the different fitting functions for $\mathcal{F}(x)$. Similar to Figure~\ref{fig1}, as $a$ increases, the intensity on the left side of the ellipse increases. Figure~\ref{fig10} shows the impact of $\varpi$ and $\varsigma$ on the shadow image. It can be observed that increasing $\varpi$ and $\varsigma$ reduces the size of the ellipse. Figures~\ref{fig11} and~\ref{fig12} show the corresponding intensity distributions. For both the x-axis and the y-axis, under anisotropic radiation, the distance between the two peaks is smaller than in the case of isotropic radiation. The above analysis indicates that the radiation type affects both the shape and size of the higher-order images.

\subsection{BAAF Model}
\begin{figure}[!h]
	\centering 
	\subfigure[$a=0.001$]{\includegraphics[scale=0.35]{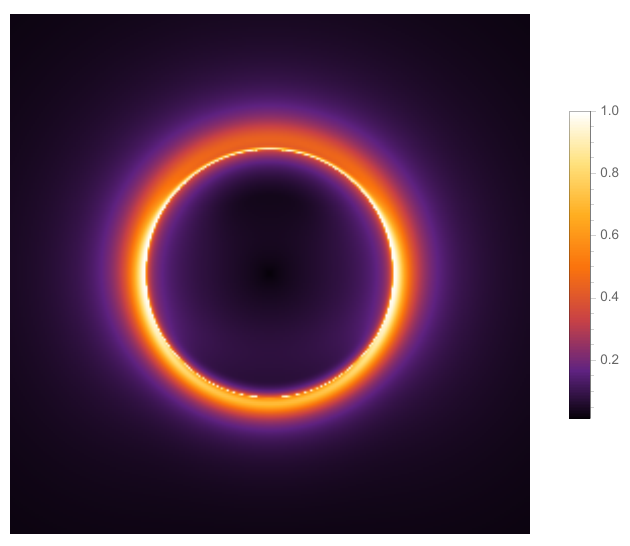}}
	\subfigure[$a=0.1$]{\includegraphics[scale=0.35]{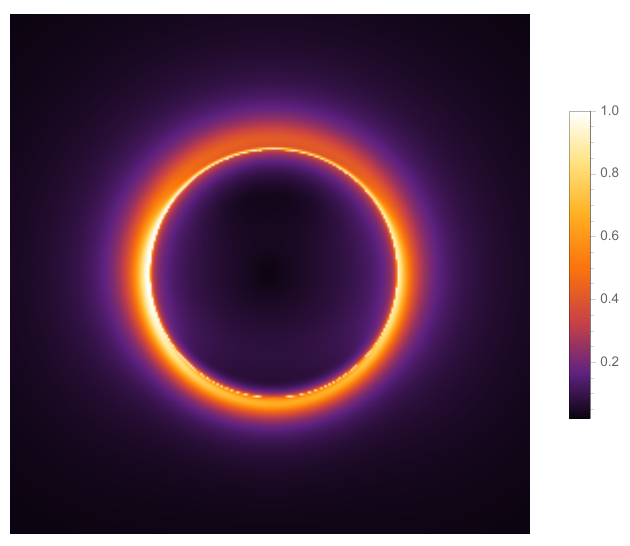}}
	\subfigure[$a=0.5$]{\includegraphics[scale=0.35]{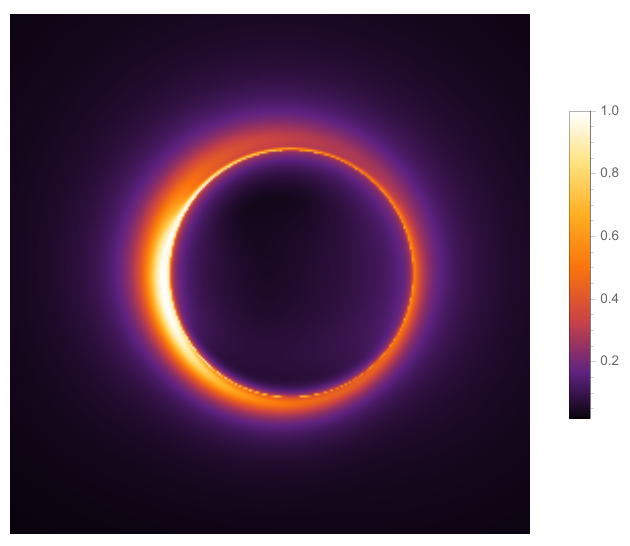}}
	\subfigure[$a=0.98$]{\includegraphics[scale=0.35]{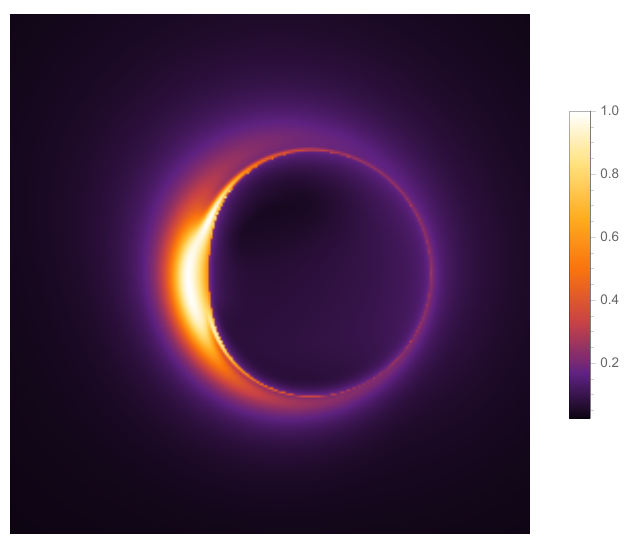}}
	
	\caption{The impact of the spin parameter $a$ on the BAAF model. The fixed parameters are    $\varpi = \varsigma = 0.1$, and $\theta_o = 65^\circ$.}
	\label{fig13}
\end{figure}

\begin{figure}[!h]
	\centering 
	\subfigure[$a=0.001$]{\includegraphics[scale=0.35]{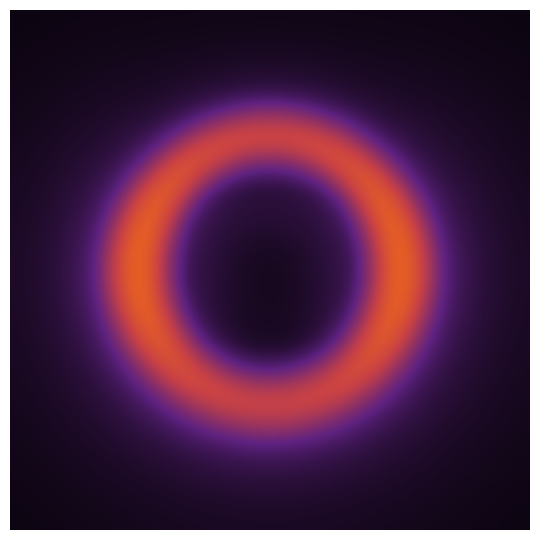}}
	\hspace{0.35cm} 
	\subfigure[$a=0.1$]{\includegraphics[scale=0.35]{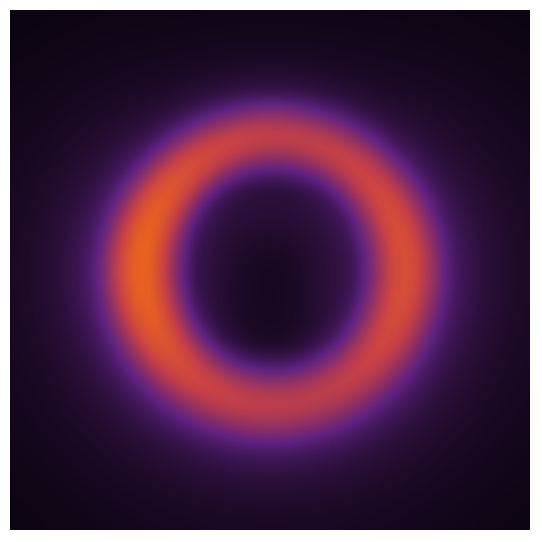}}
	\hspace{0.35cm} 
	\subfigure[$a=0.5$]{\includegraphics[scale=0.35]{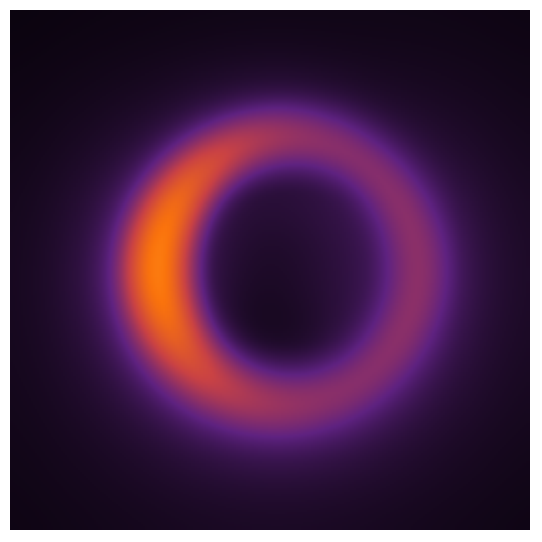}}
	\hspace{0.35cm} 
	\subfigure[$a=0.98$]{\includegraphics[scale=0.35]{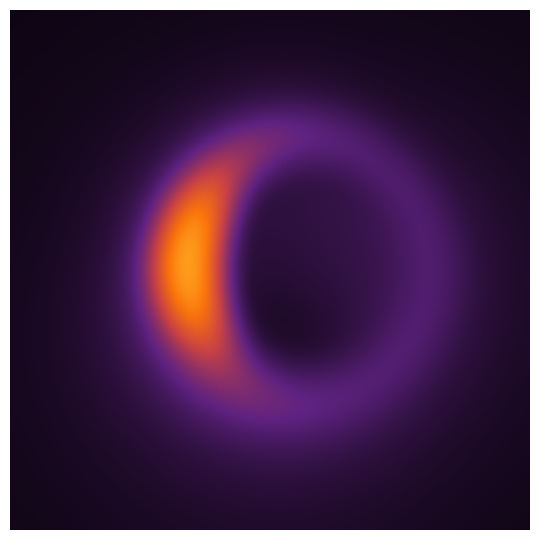}}
	
	\caption{Blurred images of black holes processed with a Gaussian filter, where the standard deviation is set to $1/12$ of the field of view $\gamma_{\mathrm{fov}}$. The plotting parameters are consistent with those in Figure~\ref{fig13}.}
	\label{fig14}
\end{figure}

\begin{figure}[!h]
	\centering 
	
	\subfigure[$\varpi=0.3,\varsigma=0.69$]{\includegraphics[scale=0.35]{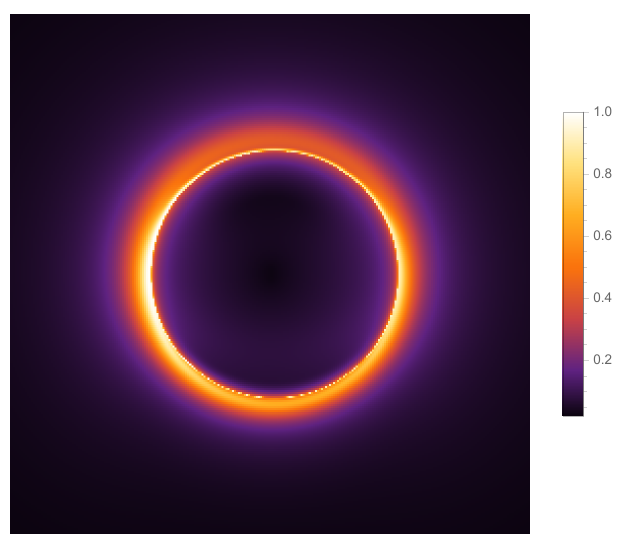}}
	\subfigure[$\varpi=0.3,\varsigma=0.79$]{\includegraphics[scale=0.35]{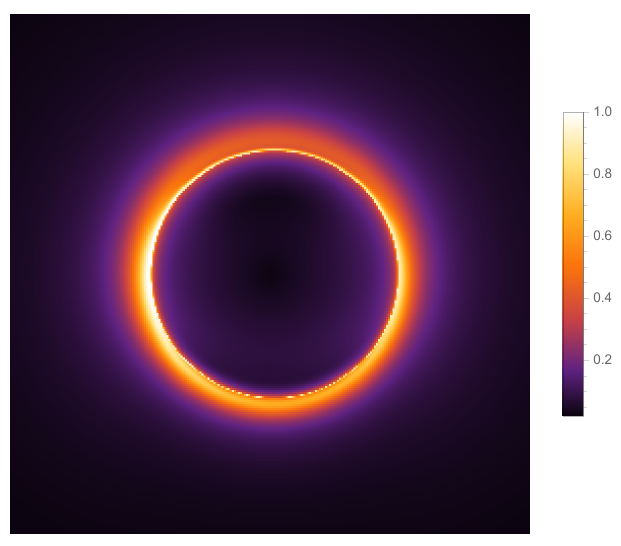}}
	\subfigure[$\varpi=0.3,\varsigma=0.89$]{\includegraphics[scale=0.35]{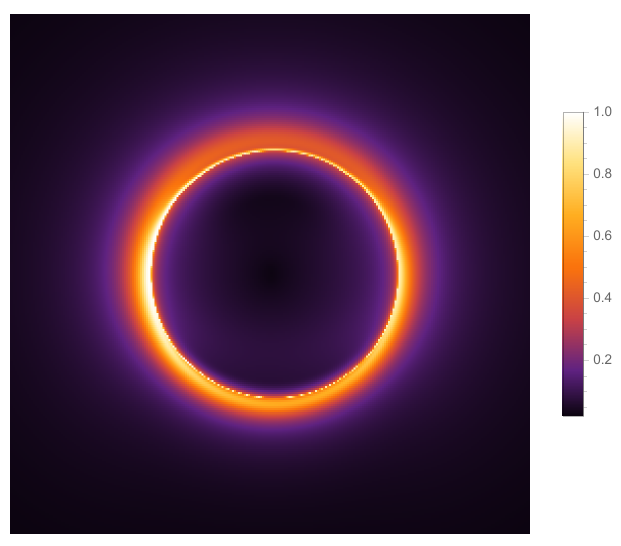}}
	\subfigure[$\varpi=0.3,\varsigma=0.99$]{\includegraphics[scale=0.35]{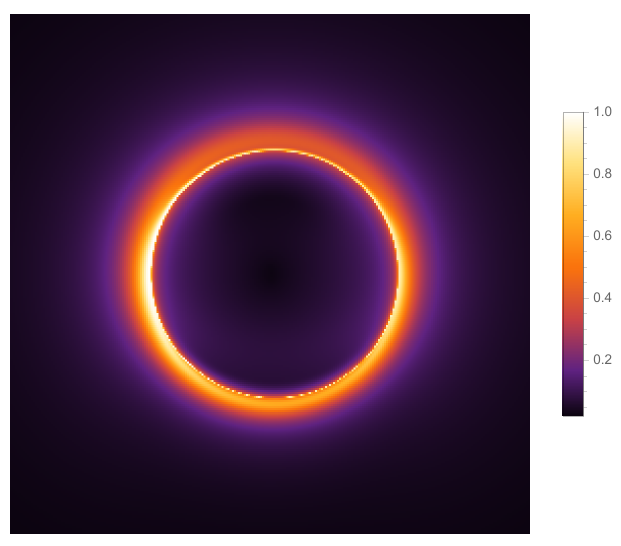}}
	
	\subfigure[$\varpi=0.6,\varsigma=0.69$]{\includegraphics[scale=0.35]{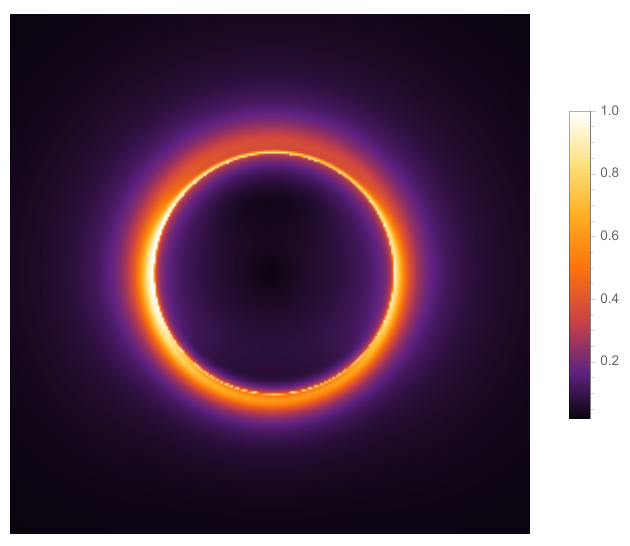}}
	\subfigure[$\varpi=0.6,\varsigma=0.79$]{\includegraphics[scale=0.35]{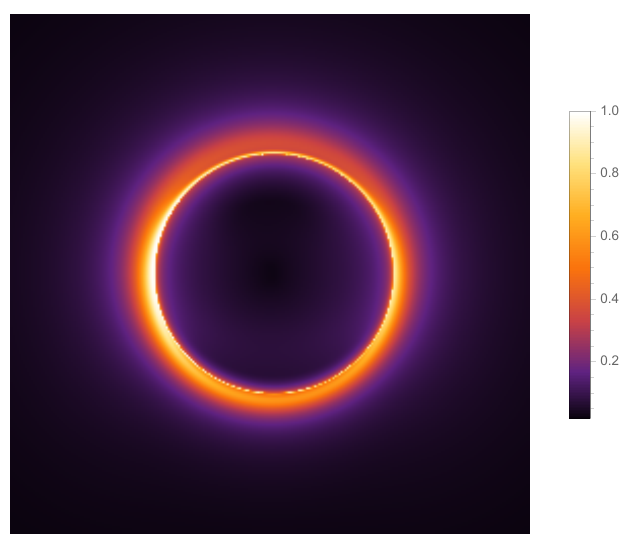}}
	\subfigure[$\varpi=0.6,\varsigma=0.89$]{\includegraphics[scale=0.35]{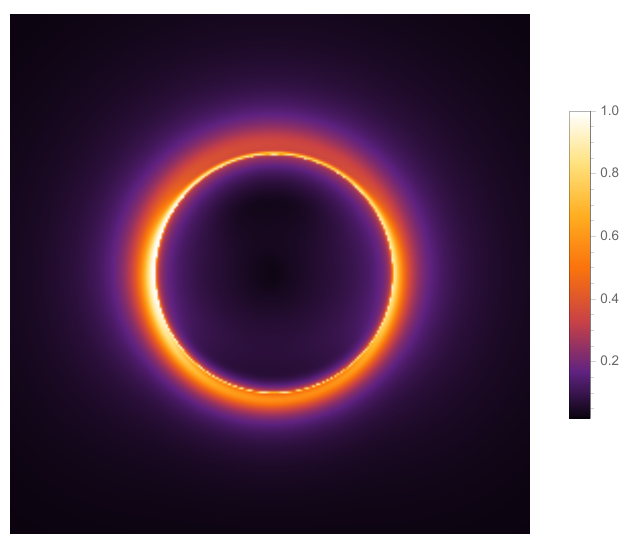}}
	\subfigure[$\varpi=0.6,\varsigma=0.99$]{\includegraphics[scale=0.35]{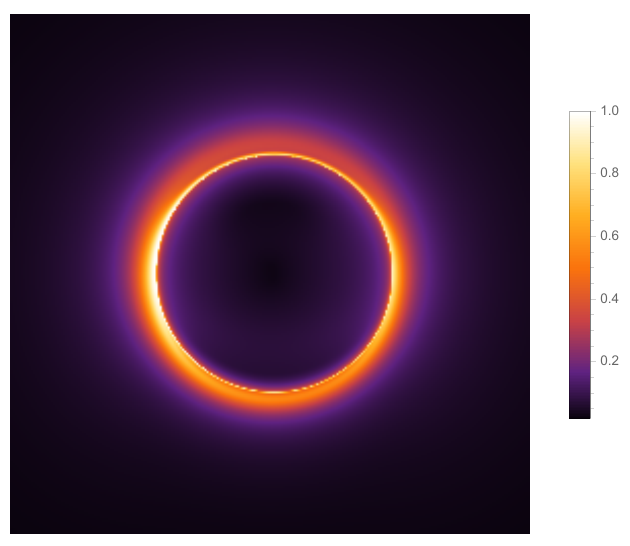}}
	
	\subfigure[$\varpi=0.9,\varsigma=0.69$]{\includegraphics[scale=0.35]{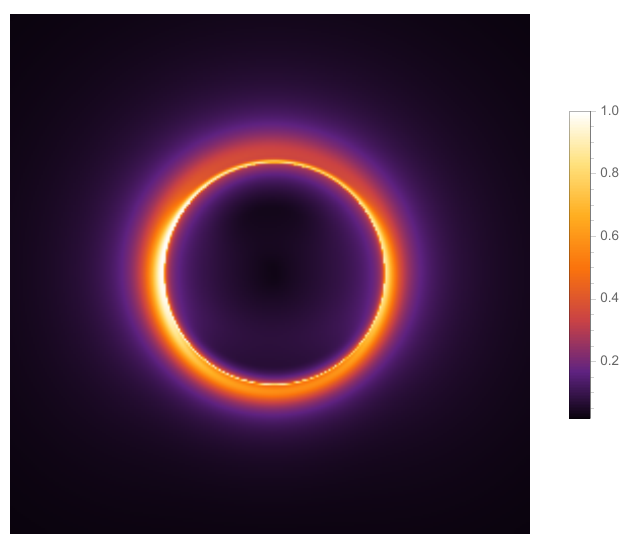}}
	\subfigure[$\varpi=0.9,\varsigma=0.79$]{\includegraphics[scale=0.35]{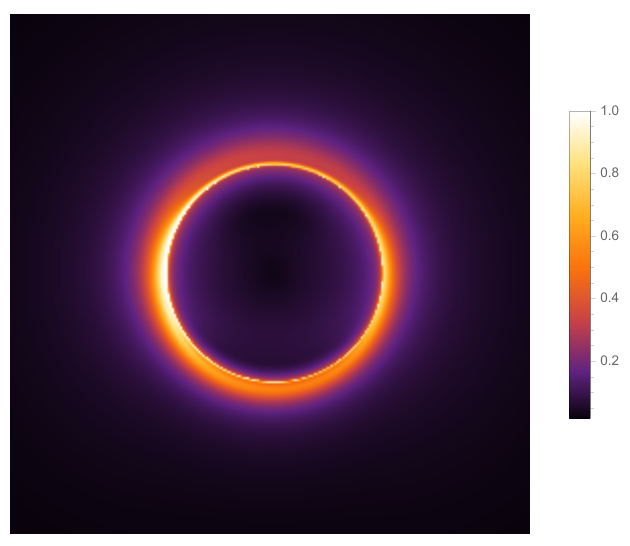}}
	\subfigure[$\varpi=0.9,\varsigma=0.89$]{\includegraphics[scale=0.35]{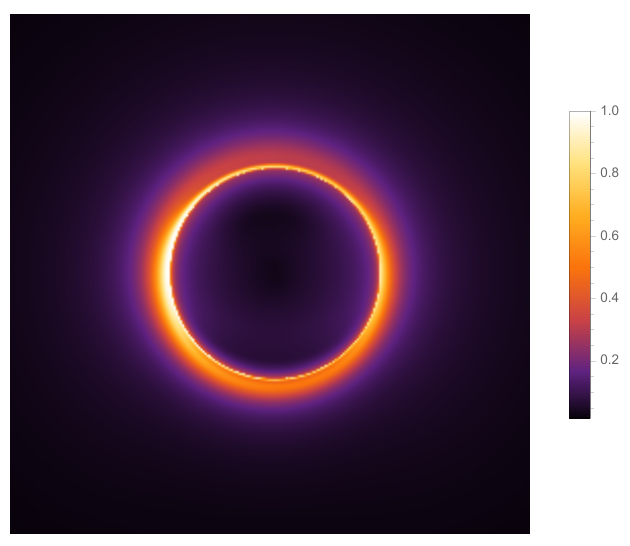}}
	\subfigure[$\varpi=0.9,\varsigma=0.99$]{\includegraphics[scale=0.35]{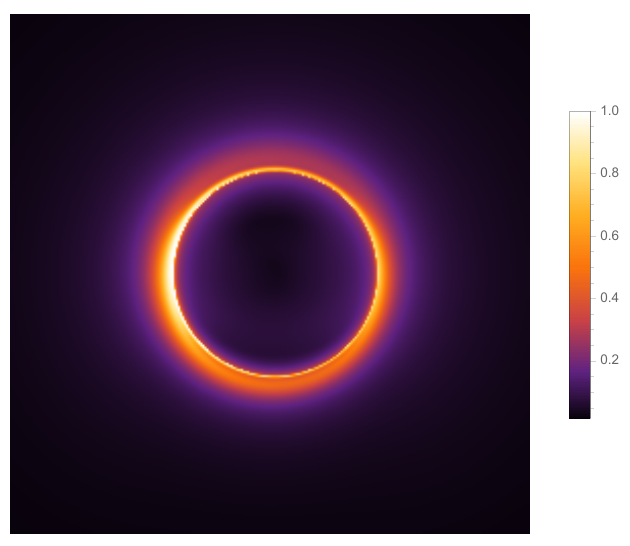}}
	
	\caption{Black hole shadow image under the BAAF model. The parameters are fixed at $a=0.1, \theta_{o}=65^\circ$.}
	\label{fig15}
\end{figure}

\begin{figure}[!h]
	\centering 
	\subfigure[the x-axis]{\includegraphics[scale=0.8]{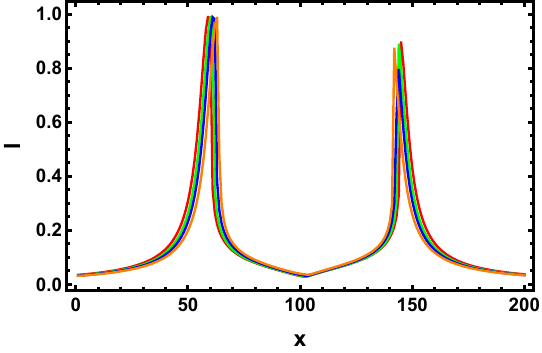}}
	\subfigure[the y-axis]{\includegraphics[scale=0.8]{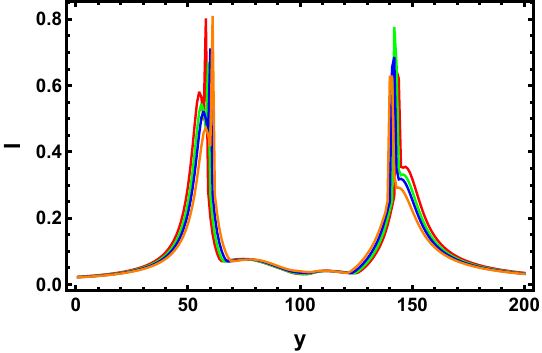}}
	
	\caption{Intensity distribution along the x-axis and y-axis for the BAAF model. The red, green, blue, and orange colors correspond to $\varsigma=0.69, 0.79, 0.89, 0.99$, respectively. The parameters are fixed as $a=0.1, \varpi=0.9, \theta=65^\circ$.}
	\label{fig16}
\end{figure}

\begin{figure}[!h]
	\centering 
	\subfigure[the x-axis]{\includegraphics[scale=0.8]{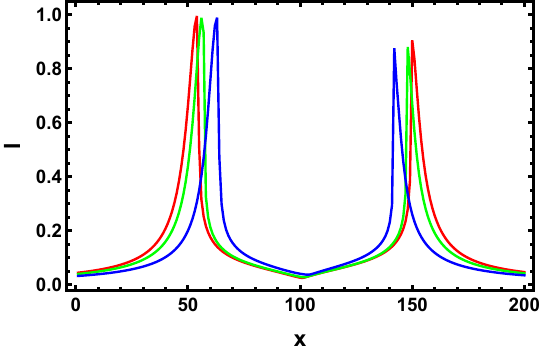}}
	\subfigure[the y-axis]{\includegraphics[scale=0.8]{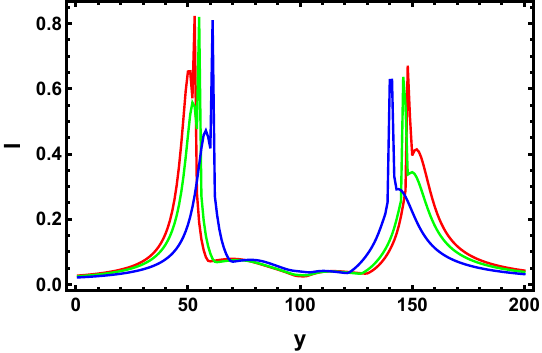}}
	
	\caption{Intensity distribution along the x-axis and y-axis for the BAAF model. The red, green, and blue colors correspond to $\varpi=0.3, 0.6, 0.9$, respectively. The parameters are fixed as $a=0.1, \varsigma=0.99, \theta_{o}=65^\circ$.}
	\label{fig17}
\end{figure}

Figure~\ref{fig13} shows the impact of the spin parameter $a$ on the BAAF model, and Figure~\ref{fig14} presents the corresponding blurred images. When $a$ is small, the higher-order images appear circular, similar to the case of isotropic radiation. However, unlike the isotropic case, the higher-order images in the BAAF model feature a complete region of intensity reduction inside them. This indicates that, compared to the RIAF model, the radiation from outside the equatorial plane in the BAAF model has a weaker effect in obscuring the event horizon. Comparing the RIAF model and the BAAF model, we find that the latter is closer to the EHT observational results. As shown in Figure~\ref{fig15}, the impact of $\varpi$ and $\varsigma$ on the higher-order images in the BAAF model is similar to that in the RIAF model. However, the region corresponding to the primary image is significantly smaller in the BAAF model. This conclusion is also evident from the intensity distribution diagrams in Figures~\ref{fig16} and~\ref{fig17}. For both the x-axis and y-axis, the peaks corresponding to the higher-order images in the BAAF model are sharper, and the intensity rapidly approaches zero beyond the peaks. This reflects that, in the BAAF model, the intensity contribution primarily originates from the higher-order images. In contrast, in the RIAF model, the intensity of the primary image outside the higher-order images remains relatively high.

\subsection{Polarization Images Illuminated by BAAF Disk}

\begin{figure}[!h]
	\centering 
	\subfigure[Stokes parameter $\tilde{I}_{o}$]{\includegraphics[scale=0.6]{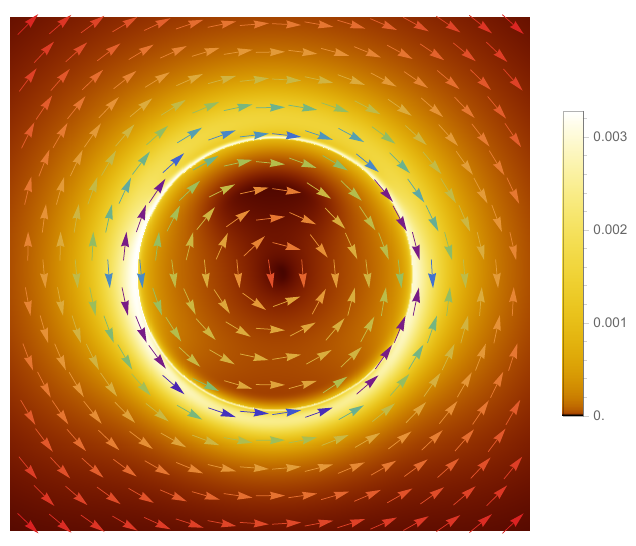}}
	\subfigure[Stokes parameter $\tilde{Q}_{o}$]{\includegraphics[scale=0.6]{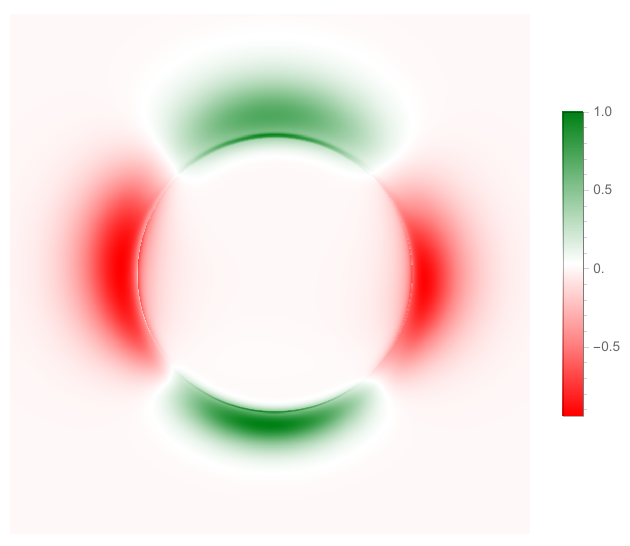}}
	\subfigure[Stokes parameter $\tilde{U}_{o}$]{\includegraphics[scale=0.6]{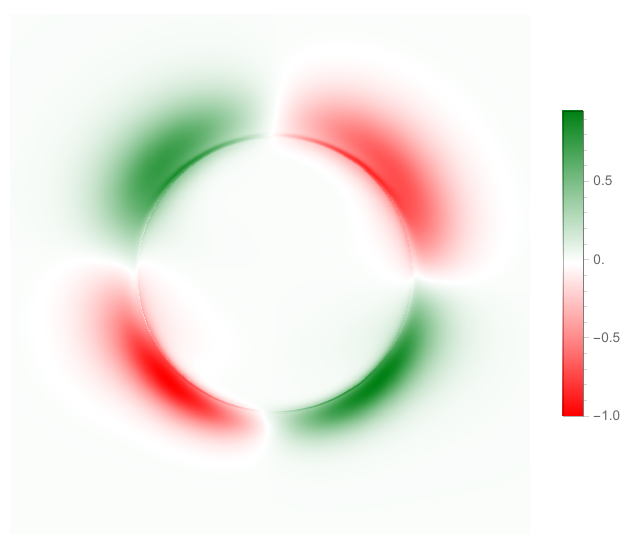}}
	\subfigure[Stokes parameter $\tilde{V}_{o}$]{\includegraphics[scale=0.6]{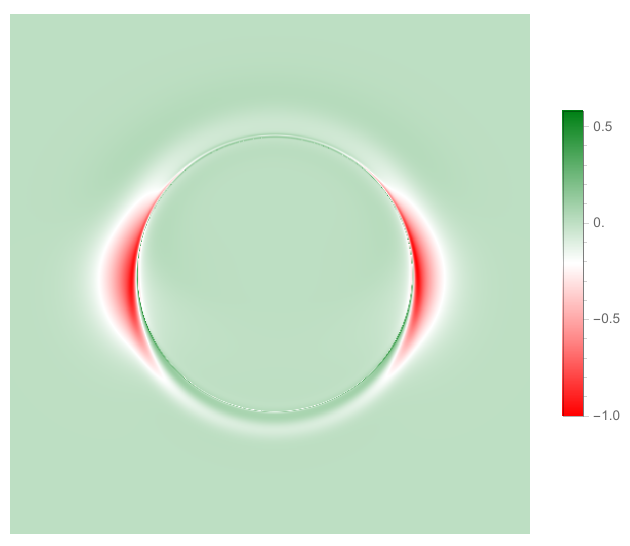}}
	
	\caption{Stokes parameters $\tilde{I}_{o}, \tilde{Q}_{o}, \tilde{U}_{o}, \tilde{V}_{o}$ in the BAAF model. The fixed parameters are $a=0.1, \varpi=0.3, \varsigma=0.69, \theta=65^\circ$.}
	\label{fig18}
\end{figure}

For the Polarization imaging, we will take the BAAF model as an example. 
Figure~\ref{fig18} shows the numerical results for the Stokes parameters $\tilde{I}_o$, $\tilde{Q}_o$, $\tilde{U}_o$, and $\tilde{V}_o$. Here, $\tilde{I}_o$ reflects the intensity distribution, with arrows representing the linear polarization vector $\vec{f}$. The color and direction of the arrows correspond to the degree of linear polarization $\mathcal{P}_o$ and the electric vector position angle $\Phi_{\mathrm{EVPA}}$. Since $\vec{f} \perp \vec{b}$, it can be inferred that the magnetic field is roughly radially distributed. The parameters $\tilde{Q}_o$ and $\tilde{U}_o$ reach their peaks near the higher-order images and decay rapidly away from this region. The $\tilde{V}_o$ parameter indicates right-handed circular polarization on both sides of the higher-order images, with left-handed circular polarization in the remaining regions.

\begin{figure}[!h]
	\centering 
	\subfigure[$\varpi=0.3,\varsigma=0.69$]{\includegraphics[scale=0.35]{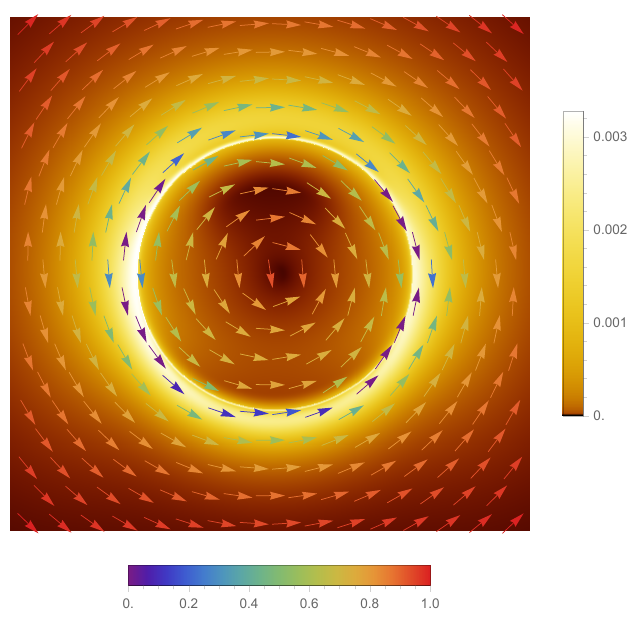}}
	\subfigure[$\varpi=0.3,\varsigma=0.79$]{\includegraphics[scale=0.35]{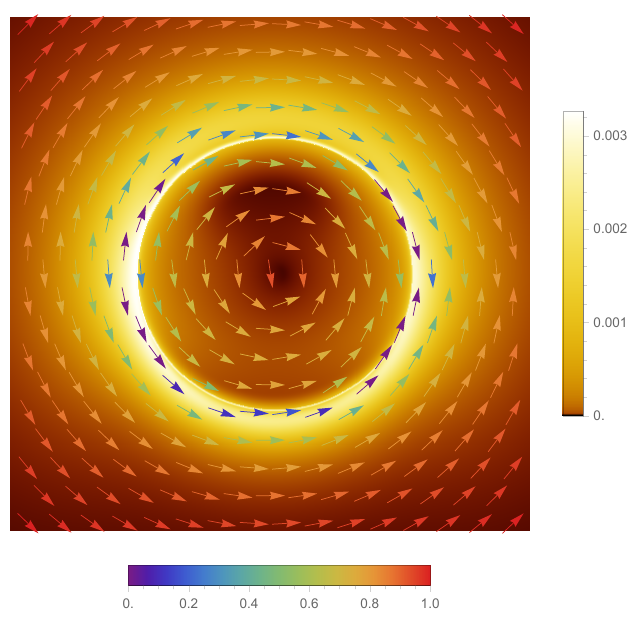}}
	\subfigure[$\varpi=0.3,\varsigma=0.89$]{\includegraphics[scale=0.35]{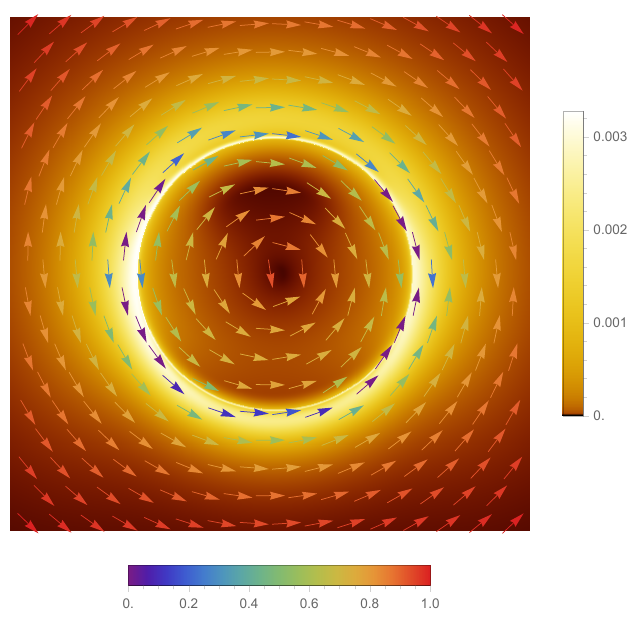}}
	\subfigure[$\varpi=0.3,\varsigma=0.99$]{\includegraphics[scale=0.35]{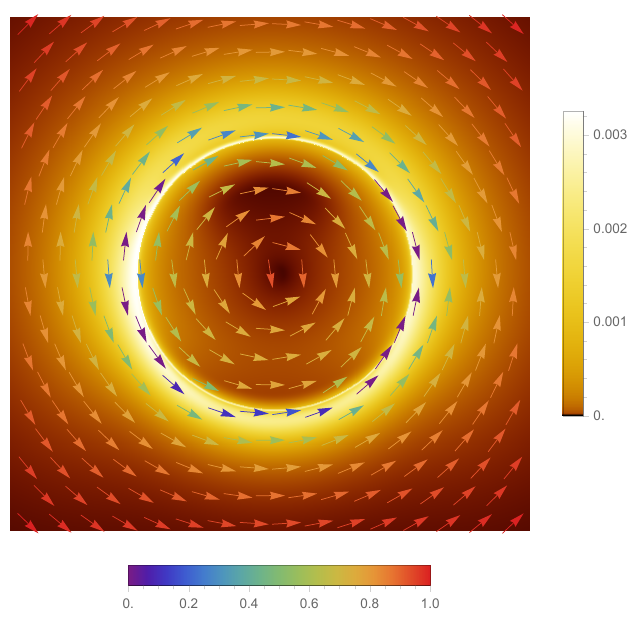}}
	
	\subfigure[$\varpi=0.6,\varsigma=0.69$]{\includegraphics[scale=0.35]{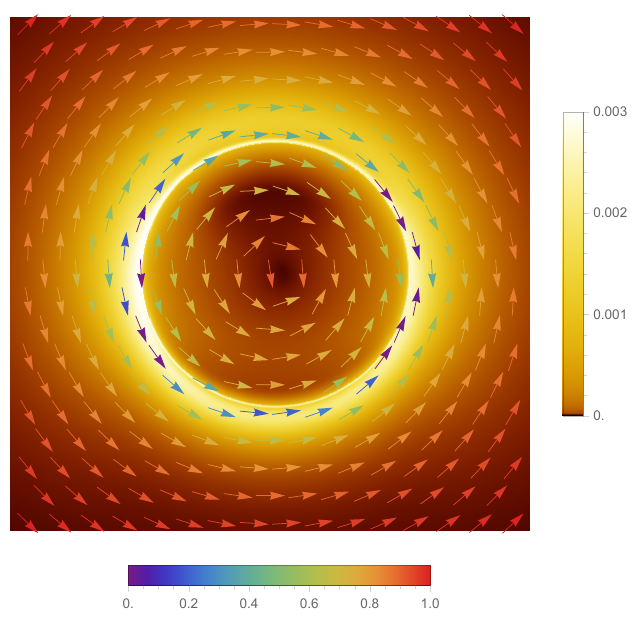}}
	\subfigure[$\varpi=0.6,\varsigma=0.79$]{\includegraphics[scale=0.35]{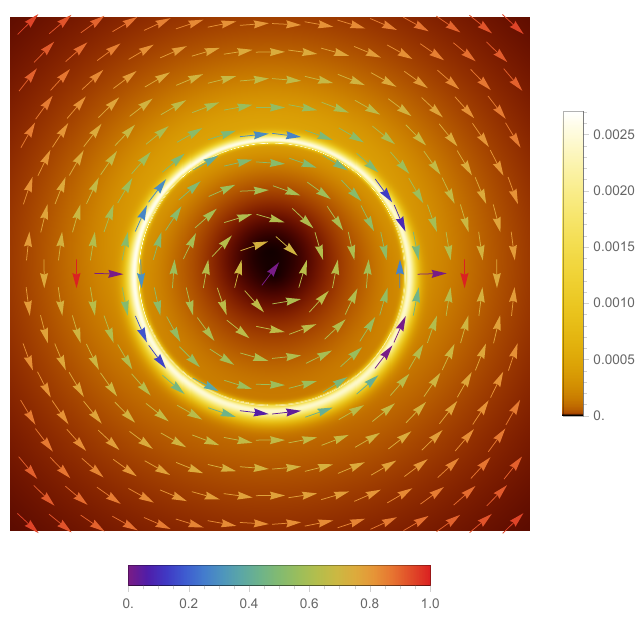}}
	\subfigure[$\varpi=0.6,\varsigma=0.89$]{\includegraphics[scale=0.35]{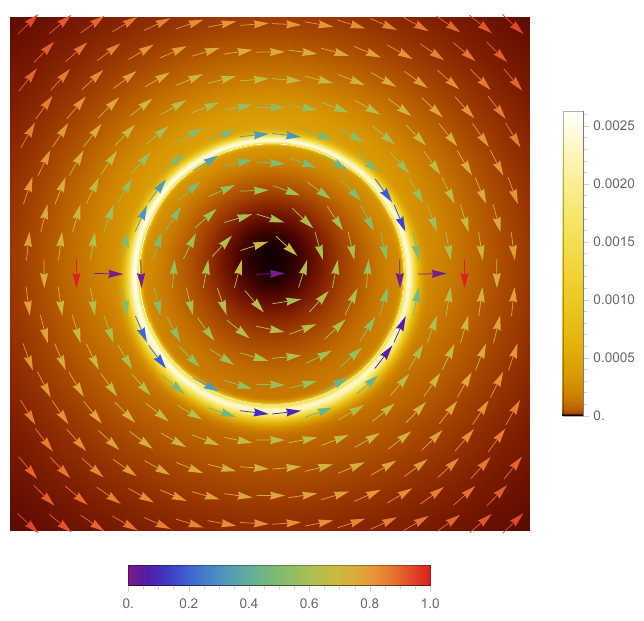}}
	\subfigure[$\varpi=0.6,\varsigma=0.99$]{\includegraphics[scale=0.35]{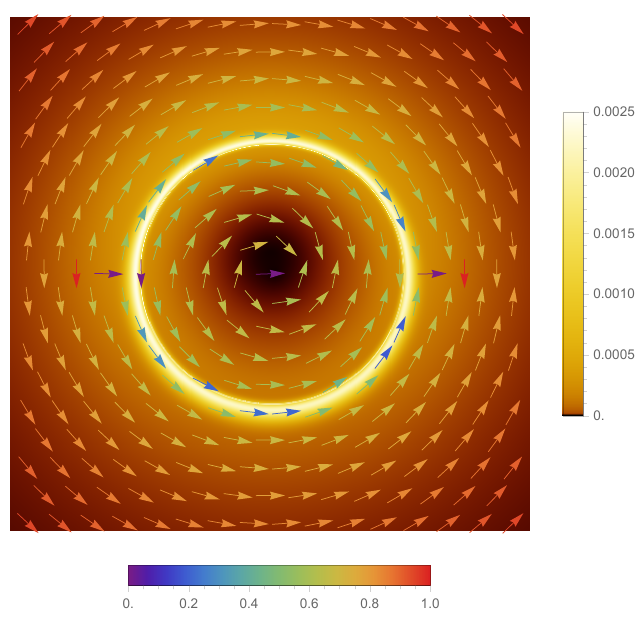}}
	
	\subfigure[$\varpi=0.9,\varsigma=0.69$]{\includegraphics[scale=0.35]{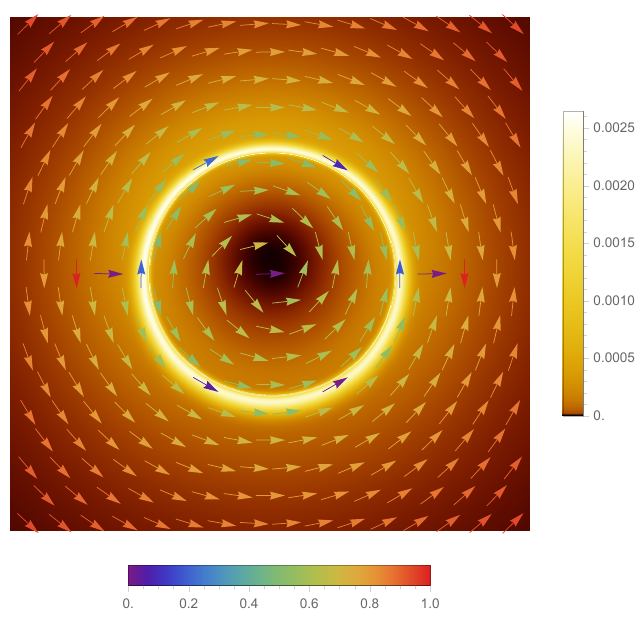}}
	\subfigure[$\varpi=0.9,\varsigma=0.79$]{\includegraphics[scale=0.35]{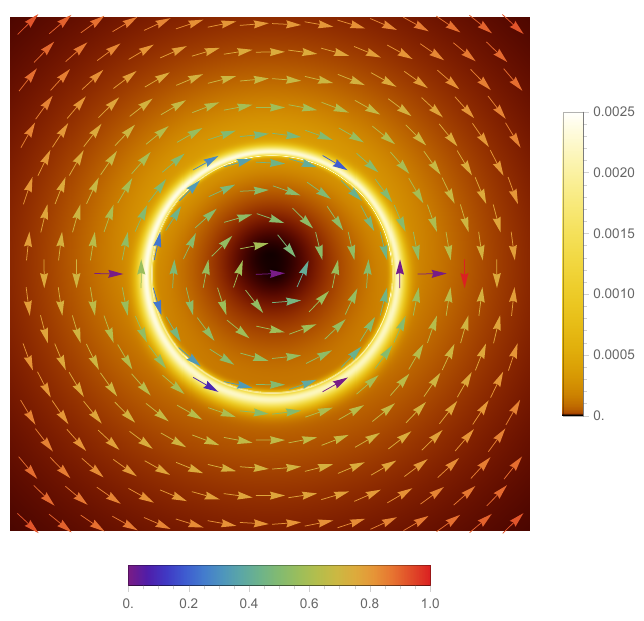}}
	\subfigure[$\varpi=0.9,\varsigma=0.89$]{\includegraphics[scale=0.35]{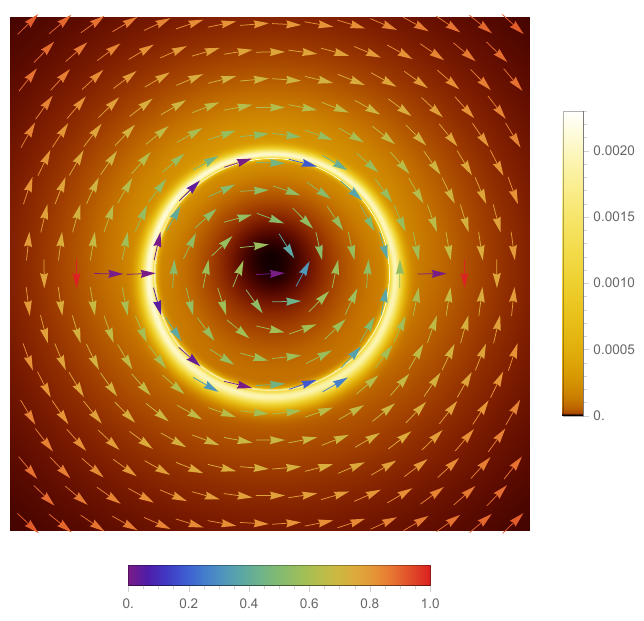}}
	\subfigure[$\varpi=0.9,\varsigma=0.99$]{\includegraphics[scale=0.35]{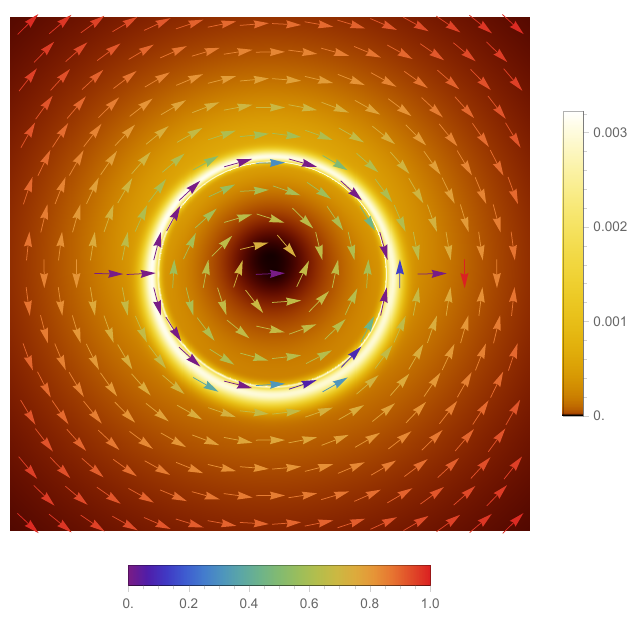}}
	
	\caption{Polarized images of black hole shadow in the BAAF model. The accretion flow follows the infalling motion, with fixed parameters $a=0.1, \theta=65^\circ$.}
	\label{fig19}
\end{figure}

Figure~\ref{fig19} shows the impact of $\varpi$ and $\varsigma$ on the polarization images. The intensity $\tilde{I}_o$ near the higher-order images increases, and from equation (\ref{eq:pv}), we observe that $\mathcal{P}_o$ decreases rapidly. Away from the higher-order images, $\mathcal{P}_o$ gradually increases. For different values of $\varpi$ and $\varsigma$, significant differences in the polarization images are observed, reflecting the impact of the spacetime’s intrinsic structure on the polarization characteristics. Interestingly, due to the presence of radiation outside the equatorial plane, polarization vectors appear across the entire imaging plane in the thick disk model, a phenomenon similar to that observed in Bardeen boson stars~\cite{li2025observational,zeng2025polarization}. However, in the thin disk model, no polarization effects are observed in the inner shadow region~\cite{yang2025observational}.

\section{Conclusion and Discussion}\label{sec7}
This paper investigates the optical and polarization imaging characteristics of rotating black holes in KR Gravity under thick accretion disk models. First, we reviewed the basic properties of KR black holes and provided definitions for null geodesics and the photon sphere. Then, we considered two thick accretion disk models: the RIAF model and the BAAF model. By numerically solving the geodesic equations and the radiative transfer equations, we obtained the corresponding black hole optical images and polarization structures.

For the RIAF model, we considered both isotropic and anisotropic radiation scenarios. In the case of isotropic radiation, we fixed the accretion flow motion to the infalling motion, with an observation frequency of $230\,\mathrm{GHz}$. The results show that increasing the spin parameter $a$ enhances the coordinate dragging effect, making the intensity on the left side of the higher-order images significantly greater than on the right. The increase in the spontaneous Lorentz violating parameters $\varsigma$ and $\varpi$ reduces the overall size of the higher-order images, but has little effect on their shape. Notably, unlike the geometrically thin accretion disk, at high observation inclination angles, the inner shadow  in the thick disk model is obscured by radiation from outside the equatorial plane, causing the central dark region to split into two parts. The impact of $(a, \varsigma, \varpi)$ on the anisotropic radiation images follows a similar trend to that of isotropic radiation, but in the anisotropic case, the higher-order images are stretched vertically, forming an elliptical shape that reflects the geometric properties of the magnetic field structure. To match the limited resolution of EHT and to more intuitively display the intensity distribution, we further plotted the blurred images and intensity distributions along the $x$–axis and $y$–axis. The results show that under Gaussian blurring, the distinguishability of the event horizon’s outline is reduced, while the intensity distribution clearly reflects the varying of the  size of  higher-order image  with model dependent  parameters.

For the BAAF model, we found that the distribution of higher-order images is narrower compared to the RIAF model, and the effect of radiation from outside of the equatorial plane on the inner shadow is significantly reduced. This is because, for specific parameters, the BAAF is geometrically thinner in some regions under the conical approximation. For polarization images, the linear polarization is negatively correlated with the intensity, and the minimum polarization degree occurs near the higher-order images. The magnitude and direction of the polarization vector are highly sensitive to $(\varsigma, \varpi)$, indicating that the polarization characteristics of rotating KR black holes can effectively reflect their intrinsic spacetime structure. Interestingly, unlike the thin accretion disk model, in the thick disk model, the gravitational lensing effect of radiation from outside of the equatorial plane causes polarization vectors to spread across the entire image plane, leading to more complex polarization patterns.

This study highlights the significant impact of black hole parameters on shadow images under thick accretion disks. Compared to geometrically thin accretion disks, the geometrically thick and optically thin accretion disk reflects more accurately the true astrophysical environment. By combining optical imaging with polarization effects, the thick disks can more comprehensively reveal the radiation characteristics and spacetime structure around black holes. Future studies will further compare black hole shadows with those of other compact objects (such as neutron stars or boson stars) to explore observable features under different gravitational sources. These efforts are expected to provide valuable theoretical insights and guidance for high-resolution astronomical observations.


\cleardoublepage

\vspace{10pt}
\noindent {\bf Acknowledgments}

\noindent
This work is supported by the National Natural Science Foundation of China (Grants Nos. 12375043,
12575069 ).
\bibliographystyle{JHEP} 
\bibliography{biblio} 

\end{document}